\documentclass[preprintnumbers, floatfix, preprintnumbers, letterpaper, superscriptaddress,nofootinbib]{revtex4}
\usepackage{graphicx}
\usepackage{microtype}
\usepackage{amsmath}
\usepackage{booktabs} 
\usepackage{siunitx}  
\usepackage{amssymb}
\usepackage{subfigure}
\usepackage{hyperref}
\usepackage{url}
\usepackage{adjustbox}
\usepackage{xcolor}
\usepackage{color}
\usepackage{mathrsfs}
\usepackage{calrsfs}
\usepackage{amsfonts}
\usepackage{lipsum}
\usepackage{eufrak}
\usepackage{tabularx}
\usepackage{eucal}
\usepackage{latexsym}
\usepackage{ragged2e}
\usepackage{epsfig}
\usepackage{textcomp}
\usepackage{float}
\usepackage{orcidlink}

\usepackage{caption}
\DeclareCaptionJustification{justified}{\leftskip=0pt \rightskip=0pt \parfillskip=0pt plus 1fil}
\definecolor{vividviolet}{rgb}{0.62, 0.0, 1.0}
\definecolor{amaranth}{rgb}{0.9, 0.17, 0.31}
\definecolor{palatinateblue}{rgb}{0.15, 0.23, 0.89}
\definecolor{brightpink}{rgb}{1.0, 0.0, 0.5}
\definecolor{cornflowerblue}{rgb}{0.39, 0.58, 0.93}
\definecolor{deepcarminepink}{rgb}{0.94, 0.19, 0.22}
\definecolor{radicalred}{rgb}{1.0, 0.21, 0.37}
\definecolor{darkyellow}{rgb}{0.7, 0.6, 0.1}

\colorlet{Mycolor1}{green!10!orange}

\hypersetup{ linktoc=all,
	colorlinks, linkcolor={palatinateblue},
	citecolor={brightpink}, urlcolor={amaranth}
}

\graphicspath{{Images/}}

\def\sideremark#1{\ifvmode\leavevmode\fi\vadjust{\vbox to0pt{\vss
			\hbox to 0pt{\hskip\hsize\hskip1em
				\vbox{\hsize1.3cm\tiny\raggedright\pretolerance10000
					\noindent #1\hfill}\hss}\vbox to8pt{\vfil}\vss}}}%
\def\beq{\begin{equation}}
\def\eeq{\end{equation}}
\newcommand{\be}{\begin{equation}}
\newcommand{\ee}{\end{equation}}
\newcommand{\ba}{\begin{eqnarray}}
\newcommand{\ea}{\end{eqnarray}}

\begin{document}

\title{Exact Integrable $\Lambda$CDM-Mimicking $f(Q)$ Cosmology: Background, Stability, and Perturbations in the First Connection Branch}


	\author{Wompherdeiki Khyllep\orcidlink{0000-0003-3930-4231}}
	\email{sjwomkhyllep@gmail.com}
	\affiliation{Department of Mathematics, St. Anthony’s College, Shillong, Meghalaya 793001, India}

    \author{Daniele Gregoris\orcidlink{0000-0002-0448-3447}}
	\email{danielegregoris@libero.it}
    \thanks{Corresponding author}
	\affiliation{Centre For Cosmology and Science Popularization (CCSP), SGT University, Gurugram, Delhi-NCR, Haryana 122505, India}
	
	\author{Saikat Chakraborty\orcidlink{0000-0002-5472-304X}}
	\email{saikat.chakraborty@nwu.ac.za,\,saikat.c@chula.ac.th}
	\affiliation{Institute of Research and Development, Duy Tan University, Da Nang 550000, Vietnam}
    \affiliation{Faculty of Natural Sciences, Duy Tan University, Da Nang 550000, Vietnam}
	\affiliation{Center for Space Research, North-West University, Potchefstroom 2520, South Africa}

    \author{Jibitesh Dutta\orcidlink{0000-0002-6097-454X}}
	\email{jibitesh@nehu.ac.in}
	\affiliation{Mathematics Division, Department of Basic Sciences and Social Sciences, North Eastern Hill University, Shillong, Meghalaya 793022, India}
	\affiliation{Visiting Associate, Inter University Centre for Astronomy and Astrophysics, Pune 411 007, India}

\begin{abstract}
We re-examine the problem of mimicking the standard cosmological model, characterized by $j=1$ (where $j$ is the cosmographic jerk parameter), within the context of the first connection branch of $f(Q)$ gravity. While this problem has previously been addressed via reconstruction techniques—yielding the analytic form $f(Q)=-2\Lambda+\alpha Q + \beta\sqrt{-Q}$ with $Q=-6H^2$—here we tackle this problem from two distinct perspectives, applying either a cosmographic closure strategy or an auxiliary-variable hierarchy approach to the traditional dynamical systems formulation. Remarkably, the cosmographic closure renders the dynamical system integrable for $\Lambda$CDM-mimicking $f(Q)$ models. Consequently, we obtain closed-form analytical solutions for all relevant cosmological quantities at both the background and linear perturbation levels, with the perturbative solutions elegantly expressed in terms of generalized Heun functions. Furthermore, we analyze the structural stability of the $\Lambda$CDM-mimicking phase space against small kinematic deviations from $j=1$, demonstrating the robustness of these solutions. Our approach provides a systematic route to studying $\Lambda$CDM-mimicking dynamics without requiring a closed-form analytic reconstruction of the underlying action. Finally, we utilize this framework to establish a direct comparison between $\Lambda$CDM-mimicking dynamics in the first connection branch of $f(Q)$ gravity and those in $f(R)$ gravity.
\end{abstract}

\maketitle

\section{Introduction}

The current standard model of cosmology, known as the $\Lambda$CDM model, has been receiving continuous scrutiny from both the observational and theoretical perspectives. Regarding the former, state-of-the-art reconstruction techniques, based on Gaussian processes applied to the Hubble rate and luminosity distance, could confirm the astrophysical relevance of such a model \cite{Jesus:2022xwb}. DESI data interpreted in light of dynamical dark energy templates do not provide conclusive evidence against the $\Lambda$CDM model \cite{Chaudhary:2025bfs}. On the other hand, DESI and Pantheon data jointly analyzed suggest a preference towards dynamical dark energy at $2\sigma$ \cite{Rodrigues:2025tfg}, possibly due to systematics in low-redshift supernova data \cite{Capozziello:2025qmh} and/or in the assumed value of the sound horizon \cite{Adi:2025hyj}. From the theoretical viewpoint, the $\Lambda$CDM model provides an elegant description of the cosmic history being structurally stable because the past dark-matter dominated and the future de Sitter epochs are both reached independently of the initial conditions \cite{Chakraborty:2022evc}. 

It is important to distinguish between quantities that are directly constrained by observations and those that are inferred within a specific cosmological model. One can use data sets such as the ones released by DESI DR2, which measure baryon acoustic oscillations and redshift-space distortions, to directly constrain the background expansion history vis-à-vis cosmographic parameters \cite{Rodrigues:2025tfg}. On the other hand, the matter abundance $\Omega_m(z)$ is not directly measured from such datasets in a model-independent manner, but is only inferred from global fits under an assumed cosmological framework, e.g., CPL \cite{DESI:2025zgx}.  Hence, different gravitational theories may reproduce essentially the same background expansion history while predicting different evolutions of the effective matter abundance and of the growth of cosmic structures \cite{Chakraborty:2021jku,Chakraborty:2022evc,Chakraborty:2025qlv}.

The $\Lambda$CDM model is kinematically characterized by the property $j(z)=1$ for the cosmographic function at every epoch, where $z$ is the cosmological redshift and $j$ is the cosmographic jerk function. In general, this condition can be viewed as a differential equation for the Hubble function $H(z)$, whose solution will consequently exhibit the same $\Lambda$CDM redshift dependence (hence preserving its structural stability and existence of a dark matter-dominated epoch in the past and a dark energy dominated in the future), but involving some arbitrary integration constants. The specific relations between these integration constants and the matter abundance parameters in the integrated expression $H(z)$ are different for different models and provide some degrees of freedom (for detailed discussion, see \cite[Sec.1]{Chakraborty:2022evc}). This way of thinking gives rise to a rich class of so-called \emph{$\Lambda$CDM-mimicking} cosmological models that share the same kinematical properties as the General Relativistic (GR) $\Lambda$CDM model at the background level, but display different evolution of perturbations due to different underlying dynamics. In fact, a derivative lacks the integration constant needed to fully encode its original function, a structural limitation shared by the Cauchy-Riemann equations when reconstructing a function from only one of its parts \cite{bressoud2011historical}. Specifically, it has already been demonstrated that a $\Lambda$CDM-like cosmic evolution can be generated not only by non-interacting pressureless dust and a cosmological constant, but also by an interacting dark sector scenario \cite{Chakraborty:2022evc}, and by modified gravitational paradigms such as $f(R)$ gravity \cite{Dunsby:2010wg,He:2012rf} and its nonminimally coupled formulation \cite{Ortiz-Banos:2021jgg}, $f(G)$ gravity \cite{Myrzakulov:2010gt}, $f(R,G)$ gravity \cite{Elizalde:2010jx}, $f(T)$ gravity \cite{Setare:2013xh}, scalar-tensor gravity \cite[Sect.VI]{Murtaza:2025gra}, and $f(Q)$ gravity \cite{Chakraborty:2025qlv}. A $\Lambda$CDM-like cosmology can also arise as a subset within a more general cosmic history, as explicitly shown in the context of $f(Q)$ gravity \cite[Sect.VA]{Dutta:2025fqw}.

Cosmography allows for a model-independent testing of the astrophysical applicability of the cosmographic condition $j(z)=1$ \cite{Capozziello:2019cav}, which is the kinematical defining property of $\Lambda$CDM-mimicking cosmological models. Cosmographic methods assume only the Copernican principle encapsulated in the geometrical symmetries of the Friedmann-Lema\^itre-Robertson-Walker (FLRW) geometry and a minimal coupling between matter and geometry, so that the geodesic motion of photons can be unambiguously defined either as a shortest path or as an autoparallel transport. Both these assumptions are respected in the theoretical set-up we will consider in this paper. Essentially, from the astrophysically collected data points in the $H(z)$ versus $z$ plane, one can compute the derivatives $H'(z)$, $H''(z)$, etc. by some discretization method and then reconstruct the redshift dependence of the deceleration, jerk, span, etc. functions numerically \cite{Capozziello:2019cav}.

Modified gravity theories provide an important framework for exploring possible departures from GR. They play a crucial role in addressing various puzzling cosmological observations. Among the diverse geometric reformulations of gravity, one well-studied modified gravitational theory is symmetric teleparallel gravity. This theory is based on the nonmetricity of spacetime, characterized by the nonmetricity scalar $Q$, rather than on curvature or torsion. Its extension to $f(Q)$ gravity has attracted considerable attention and enriches the theory at both cosmological and astrophysical scales. There are three possible symmetric teleparallel connection branches respecting spatial flatness, spatial homogeneity, and isotropy, and we will confine the present work to the first connection branch, which we term $\Gamma_1$. In this branch, the field equations become second-order, thereby avoiding Ostrogradsky instabilities while still allowing a rich cosmological phenomenology. The mathematical simplicity of this branch,  makes it particularly attractive for further studies. During the last few years, $f(Q)$ gravity has been extensively investigated in various areas such as late-time cosmology \cite{BeltranJimenez:2019tme,Frusciante:2021sio}, the early inflationary Universe \cite{Capozziello:2022tvv,Capozziello:2024lsz}, black hole solutions \cite{Nashed:2025bxv}, compact stars \cite{Lin:2021uqa}, and gravitational waves \cite{Capozziello:2024vix}. Several $f(Q)$ models have also been shown to alleviate the $H_0$ and $\sigma_8$ tensions \cite{Boiza:2025xpn,Li:2025msm}. In the $\Gamma_1$ branch of $f(Q)$ gravity, 
the background dynamics of $f(Q)$ gravity is mathematically equivalent to that of $f(T)$ teleparallel gravity \cite{Jarv:2018bgs, Hohmann:2017jao}. This background equivalence reflects a deeper structural analogy. For instance, the non-metricity scalar and its conjugate in $f(Q)$ gravity map directly to their torsional counterparts in $f(T)$ gravity \cite{Capozziello:2026rrc,Paliathanasis:2025jlw}. In the specific scenario of a spatially flat spacetime, as that considered in what follows, choosing the affine connection components to respect the background cosmological symmetries   in $f(Q)$ cosmology yields $T = Q$, recovering the same set of cosmological field equations as $f(T)$ gravity \cite{DAmbrosio:2021pnd}. 

The form of the $\Lambda$CDM-mimicking $f(Q)$ gravity in the first branch of connection has been reconstructed in \cite{Chakraborty:2025qlv}; the most generic such functional form comes out as $f(Q)=-2\Lambda+\alpha Q+\beta\sqrt{-Q}$. Such a form of $f(Q)$ (modulo the cosmological constant term) has recently been considered also in \cite{Li:2025msm} because its linear perturbation analysis is well defined on a general FLRW background and ghost stability is preserved by the physically relevant values of the free parameters according to the criteria \cite{BeltranJimenez:2019tme,Hu:2023juh,Zhao:2024kri}. We also report that other functional forms, such that including a contribution from the inverse of the non-metricity, are still subject of cosmological data testing \cite{Atayde:2026upv}. 

The mathematical tool that we extensively employ in our study is the dynamical systems framework in cosmology. Nowadays, dynamical-system techniques have become indispensable tools for investigating the global evolution of cosmological models \cite{Bahamonde:2017ize}. They provide a rigorous framework for analyzing the cosmological viability and phase-space structure of complicated modified gravity theories without requiring numerical integration for individual initial conditions. This qualitative perspective perfectly complements the cosmographic approach in assessing the structural robustness of the underlying models. In fact, exact analytical solutions remain comparatively rare, especially in modified gravity theories. Therefore, dynamical-system methods allow for an understanding of the main properties of the system when  exact analytical integrations cannot be achieved.

An important question motivating the present work is whether an observed $\Lambda$CDM expansion history uniquely identifies the underlying theory of gravity. It is known that  while different gravitational theories can reproduce the same background kinematics, they predict different effective matter abundances, gravitational couplings, and growth histories of cosmic structures. The main novelty of our work   is that we develop a fully analytical framework to investigate this degeneracy within the $\Gamma_1$ branch of $f(Q)$ gravity. In particular, we study the reconstructed $\Lambda$CDM-mimicking theory  using two distinct autonomous closure schemes. It allows us to derive exact closed-form solutions for the background and linear-perturbation sectors. Further we could establish analytical relations among cosmographic variables, dynamical-system variables, and observational growth quantities.

The physical interpretation of these two routes is the following: both procedures track the evolution of the matter abundance, however, with respect to different physical aspects of the cosmological evolution. The theory-driven closure describes the evolution in terms of departures of the underlying gravitational theory from GR,  while the cosmography-driven closure organizes the evolution into stages of cosmic expansion, as characterized by the deceleration parameter $q$.

In our specific $f(Q)$ framework, having an explicit reconstructed theory allows us to perform a comparative analysis of the two closure routes, explicitly verifying their consistency, which is not possible in $f(R)$ gravity \cite{Chakraborty:2021jku}. Our  analytical approach will further allow us to provide cosmologically transparent interpretations to the mathematical parameters entering our considered $f(Q)$ functional in terms of cosmological parameters,  like the present-day values of the deceleration parameter and matter abundance.  We also provide the first explicit application of the hierarchy approach introduced in \cite{Dutta:2025fqw} to the closure of the dynamical system with a  given $f(Q)$ theory. By introducing a finite tower of auxiliary functions, this approach avoids the multi-branch ambiguities that can arise in the direct inversion of the theory. It also  provides a systematic treatment of the branch structure of the given  model.

The astrophysical inference of the jerk function is subject to uncertainties, which challenge a definitive $\Lambda$CDM-like interpretation. We therefore also provide exact analytical closed-form results for the relevant cosmological observables for an almost $\Lambda$CDM-like cosmology characterized by an evolving jerk function that slowly deviates from its unity value. In this context, we will show how closing the system via the cosmographic approach will save us from the mathematical difficulties of reconstructing the underlying gravitational theory in closed-form, but still allowing us to present the astrophysically relevant quantities and to prove the robustness (that is the structural stability of the phase portrait) of the $\Lambda$CDM-like model in the $\Gamma_1$ branch of our given theory.  Furthermore, we will demonstrate the stability of the $\Lambda$CDM-like model by including small deviations of the jerk function from unity, under homogeneous and isotropic perturbations.

As previously mentioned, degeneracies between different cosmological models at the background level can be broken by studying the evolution of their matter density contrast. To connect this perturbation analysis with our cosmographic closure at the background level, we express the differential equation governing the relative variation of the matter density contrast in terms of the deceleration parameter. This allows us to obtain an analytical closed-form solution in terms of Heun functions. Remarkably, our analysis also shows that the growth-index parameter approaches its $\Lambda$CDM value towards the present cosmic epoch.

Our paper is organized as follows. We begin by presenting, for completeness, the Friedmann and Raychaudhuri equations in a spatially flat FLRW universe in $f(Q)$ gravity and by recalling the problems of closing their dynamical-system formulations for generic functionals $f(Q)$ in Sects. \ref{sec:f(Q)_cosmology} and \ref{sec:DSframework}, respectively. Then, we present our results for the ``closing with the theory'' and the ``closing with the cosmography'' in the two separate Sects. \ref{ss2a} and \ref{ss2b}. While the previous sections are purely mathematical, we explain the astrophysical significance of the model parameters entering the $\Lambda$CDM-mimicking function $f(Q)$ in Sect. \ref{ss2c}. We establish the robustness and stability of our $\Lambda$CDM-mimicking $f(Q)$ model in Sects. \ref{sec:robustness} and \ref{sec:stability}, respectively, and study the evolution of the matter perturbations in Sect. \ref{sec:ptbn}.  We conclude in Sect. with a comparative analysis of how our gravity model accounts for $\Lambda$CDM kinematics relative to other modified gravity paradigms, and by putting our results in the perspective of current cosmological observational tensions.

\section{$f(Q)$-gravity field equations in a spatially flat Friedmann universe}\label{sec:f(Q)_cosmology}

In the framework of symmetric teleparallel gravity, spacetime is characterized by vanishing curvature,
$R^\alpha_{\;\;\beta\mu\nu}=0$, and vanishing torsion,
$T^\alpha_{\;\;\mu\nu}=0$, while the gravitational interaction is entirely encoded in the nonmetricity tensor
\begin{equation}
Q_{\alpha\mu\nu}=\nabla_\alpha g_{\mu\nu}.
\end{equation}
We refer to the reviews \cite[Sect.2.1.5]{Bahamonde:2021gfp} and \cite[Sect.3E]{Heisenberg:2023lru} for locating the symmetric teleparallel gravity inside the more general landscape of metric-affine theories.  The $f(Q)$ theory is obtained by promoting the nonmetricity scalar $Q$ in the gravitational action to an arbitrary function $f(Q)$.

 We consider a spatially flat homogeneous and isotropic FLRW universe with scale factor $a(t)$ and Hubble parameter
\begin{equation}
H=\frac{\dot a}{a},
\end{equation}
where an overdot here and throughout the text denotes differentiation with respect to cosmic time.

In homogeneous and isotropic cosmology, several connection branches 
are compatible with the FLRW geometry. In the present work, we 
restrict our attention to the $\Gamma_1$ branch, 
for which the affine connection vanishes and the nonmetricity scalar 
acquires the particularly simple form
\begin{equation}
Q=-6H^2.
\label{Q-1}
\end{equation}

We choose this  branch as it  constitutes the simplest realization of cosmological 
$f(Q)$ gravity and provides a suitable framework for comparing 
different dynamical closure prescriptions.

The matter sector is assumed to consist of pressureless dust with energy density $\rho_m$, obeying the standard conservation equation
\begin{equation}
\dot{\rho}_m+3H\rho_m=0,
\end{equation}
consistent with the application of the Bianchi identities to the field equations \cite{BeltranJimenez:2017tkd}, and whose solution is given by
\begin{equation}
\rho_m=\rho_{m0}\left(\frac{a_0}{a}\right)^3.
\end{equation}

The Friedmann and Raychaudhuri equations of $f(Q)$ gravity in the $\Gamma_1$ branch are given by
\begin{subequations}
\label{fieldG1}
\begin{align}
3H^2f_Q+\frac12\left(f-Qf_Q\right)&=\rho_m,
\label{friedG1}
\\
-2\frac{d}{dt}\left(Hf_Q\right)-3H^2f_Q
-\frac12\left(f-Qf_Q\right)&=0,
\label{rayG1}
\end{align}
\end{subequations}
where a subscript $Q$ denotes differentiation with respect to the nonmetricity scalar $Q$.

The Friedmann equation can be rewritten as

\begin{equation}
3H^2=\kappa_{\rm eff}(\rho_m+\rho_Q),
\end{equation}

where

\begin{equation}\label{kappaeff}
    \kappa_{\rm eff}=\frac1{f_Q},
\end{equation}
and 
\begin{equation}
\rho_Q=\frac12(Qf_Q-f).
\end{equation}

Similarly, the Raychaudhuri equation assumes the form

\begin{equation}
-2\dot H-3H^2=\kappa_{\rm eff}p_Q,
\end{equation}

with

\begin{equation}
p_Q=-\rho_Q+2H\dot f_Q.
\end{equation}

\section{Dynamical Systems Framework for $\Lambda$CDM-Mimicking Cosmology}
\label{sec:DSframework}

In this section we establish the dynamical systems framework that will be employed throughout the paper to investigate $\Lambda$CDM-mimicking cosmologies in the $\Gamma_1$ branch of $f(Q)$ gravity. Our goal is to formulate the cosmological dynamics in terms of suitable dimensionless variables and to compare two distinct closure prescriptions. The first closure is theory-driven, relying on the explicit reconstructed form of the underlying gravitational Lagrangian, while the second is cosmography-driven and exploits purely kinematical information associated with the cosmic expansion history. Establishing a common dynamical framework allows us to compare these two approaches on equal footing.

\subsection{Dynamical variables and auxiliary functions}

Following \cite{Dutta:2025fqw}, we introduce the dimensionless dynamical variables\footnote{For the $\Gamma_1$ branch of interest of this paper, $x_2\equiv 1$ \cite{Dutta:2025fqw}.}
\begin{equation}
x_1=-\frac{f}{6H^2f_Q},
\qquad
x_3=-\frac{\dot f_Q}{Hf_Q},
\qquad
\Omega=\frac{\rho_m}{3H^2f_Q},
\label{dynvar_gamma1}
\end{equation}
together with the auxiliary functions
\begin{equation}
\label{def_r_m}
r(Q)\equiv Q\frac{f_Q}{f}
=\frac{1}{x_1},
\qquad
m(Q)\equiv Q\frac{f_{QQ}}{f_Q}.
\end{equation}

The variables $r$ and $m$ play a central role in the dynamical systems analysis. In particular, the quantity $m$ characterizes departures from the GR limit, while $r$ provides a convenient parametrization of the underlying gravitational Lagrangian.
The modified Friedmann and the Raychaudhuri equations \eqref{friedG1} and \eqref{rayG1} can be expressed as
\begin{subequations}
\ba
x_1+\Omega &=& 2\,\label{fried_G1}\,,
\\
x_3 &=& 2m(1+q)\,\label{raych_G1}.
\ea
\end{subequations}

where $q=-1-\dot H/H^2$ denotes the deceleration parameter.
The physical matter abundance parameter is related to the dynamical variables $\Omega$ as $\Omega_m=f_Q \Omega=\Omega/\kappa_{\rm eff}$.

At this stage the system is not yet closed, since the evolution equations depend on the functional form of $m(Q)$. In the remainder of this paper we shall consider two different closure prescriptions:

\begin{enumerate}
\item a \emph{theory-driven closure}, obtained from the explicit reconstructed form of the $f(Q)$ Lagrangian;

\item a \emph{cosmography-driven closure}, obtained directly from the kinematics of a $\Lambda$CDM-mimicking cosmic expansion.
\end{enumerate}

The comparison between these two closure schemes constitutes one of the main objectives of the present work.

\section{Theory-Driven Closure}
\label{ss2a}

The dynamical system for FLRW cosmology in the first connection branch of $f(Q)$ is one-dimensional. Detailed studies of the same have been carried out in \cite{Boehmer:2022wln,Dutta:2025fqw}, and we refer the reader to these references for a detailed exposition. In this paper, for our specific purpose, we merely import the relevant equations from our earlier paper \cite{Dutta:2025fqw}.

The dynamical system is constituted by the evolution equation
\beq\label{DS_CT_G1}
\Omega' = 3\Omega \left[ \Omega \left(\frac{1+m(Q)}{1+2m(Q)} \right)-1 \right]\,.
\eeq 

It has been derived, via analytic reconstruction method, in \cite{Chakraborty:2025qlv}, that the particular $f(Q)$ theory that gives rise to a $\Lambda$CDM-like cosmological evolution in the first connection branch is
\beq
\label{flcdmG1}
f(Q)=-2\Lambda +\alpha Q + \beta \sqrt{-Q}\,, \qquad \alpha,\,\beta \neq 0,  \,\,\,\,\, \Lambda >0\,.
\eeq
For this theory, the function $r(Q)$ is
\beq\label{rvsQG1}
r(Q) = \frac{\alpha Q + \frac{\beta}{2}\sqrt{-Q}}{-2\Lambda + \alpha Q + \beta\sqrt{-Q}}\,.
\eeq
The auxiliary variable $m(Q)$, when calculated in terms of $r$, becomes the following two-branch function
\ba
\label{mvsrG1}
m(Q) = \frac{\beta}{2\beta-4\alpha\sqrt{-Q}} \Longrightarrow m(r)=\frac{\beta (r-1)}{\beta \pm \sqrt{[\beta(2r-1)]^2-32\alpha\Lambda (r-1)r}}\,.
\ea
The Friedmann constraint \eqref{fried_G1} and the definition of $r$ in Eq.\eqref{def_r_m} allows us to write
\beq
\label{condr2}
r=\frac{1}{2-\Omega}\,.
\eeq
When the above expression is substituted into Eq.\eqref{mvsrG1}, one obtains a double-valued function $m(\Omega)$. This gives rise to a two-branched dynamical system from Eq.\eqref{DS_CT_G1}. 

The particular theory \eqref{flcdmG1}, without the explicit cosmological constant term, falls within the class of theories studied in Ref.\cite{Dutta:2025fqw}. However, it is precisely the appearance of an explicit cosmological constant term that makes $m(r)$ double-valued, thus complicating the subsequent dynamical system analysis.

In general, the conditions  $f_Q>0$ (which amounts to requiring that gravity is attractive) and $Q<0$ can, in principle, be exploited for setting further constraints on the physically viable region of the phase in general \cite{Murtaza:2025gme}. However, as it turns out that for this particular theory, the mathematical complexity does not allow one to obtain any further phase space constraints, as already noticed in\footnote{Note the different notations adopted for the dynamical variables.} \cite[Eq.(56)]{Murtaza:2025gme}.

\subsection{The problem with the standard dynamical system formulation: the branching issue}
	\label{LambdaG1}

We have that the discriminant
		\begin{equation}
			\label{discriminant}
			\Delta = [\beta(2r-1)]^{2}-32\alpha\Lambda(r-1)r\,,
		\end{equation}
		in (\ref{mvsrG1}) determines the number of admissible branches of the reconstructed function $m(r)$.
		
		When $\Delta>0$, the function $m(r)$ is double-valued, having two distinct real solutions; when $\Delta<0$, the two branches become complex, meaning that no real branch of $m(r)$ exists at that value of~$r$.
		
		The critical configuration $\Delta=0$ defines the characteristic value of the cosmological constant,
		\begin{equation}
			\label{lambdac}
			\Lambda_{c}(r)=\frac{[\beta(2r-1)]^{2}}{32\alpha(r-1)r},
		\end{equation}
		at which the two real branches of $m(r)$ merge continuously.
		
		The cosmological constant $\Lambda$ therefore acts as a bifurcation control parameter in the theory space spanned by $(r,m)$, causing a qualitative change in the number of solutions as it passes this critical value.
		
		Since $\Lambda>0$, the critical value $\Lambda_{c}$ is physically relevant only when it is positive, which occurs when $\alpha$ and the product $(r-1)r$ have the same sign. If they have opposite signs, then $\Lambda_{c}<0$, i.e. $\Delta>0$ for all physical values of $\Lambda$, and the system always remains in the two-solution regime.
		
		Relations between $\Lambda$ and $\Lambda_{c}$ that determine the number of branches depend on the combined signature of $\alpha\cdot(r-1)r$, as summarized in the table \ref{tab:lambda_branches}. The table shows that the inequality required for $\Delta>0$ reverses depending on whether $\alpha\cdot(r-1)r$ is positive or negative. 
\begin{table}[H]
    \centering
    \begin{tabular}{@{}l| c| c| c| c@{}}
        \toprule
        \textbf{Condition on $r$} & \textbf{Sign of $(r-1)r$} & \textbf{Sign of $\alpha$} & \textbf{Condition for $\Delta > 0$} & \textbf{Condition for $\Delta < 0$} \\
        & & & (Two Real Branches) & (No Real Branches) \\
        \midrule
        $r>1$ or $r<0$ & Positive & Positive & $\Lambda < \Lambda_{c}(r)$ & $\Lambda > \Lambda_{c}(r)$ \\
        $r>1$ or $r<0$ & Positive & Negative & $\Lambda > \Lambda_{c}(r)$ & $\Lambda < \Lambda_{c}(r)$ \\
        $0<r<1$ & Negative & Positive & $\Lambda > \Lambda_{c}(r)$ & $\Lambda < \Lambda_{c}(r)$ \\
        $0<r<1$ & Negative & Negative & $\Lambda < \Lambda_{c}(r)$ & $\Lambda > \Lambda_{c}(r)$ \\
        \bottomrule
    \end{tabular}
     \caption{Conditions for the existence of real branches of the reconstructed function $m(r)$ based on the signs of $\alpha$ and the radial parameter $r$. The bifurcation threshold is defined by the critical cosmological constant $\Lambda_c$.}
    \label{tab:lambda_branches}
\end{table}

Table~\ref{tab:lambda_branches} provides the mathematical classification of the reconstructed solutions. Cosmological evolution, however, is further constrained by the Friedmann constraint. Since
$
\Omega=\frac{2r-1}{r}\geq 0,
$
one must have
$
r<0 \quad \text{or} \quad r\geq 1/2.
$
As a result, even though Table~\ref{tab:lambda_branches} lists all the \emph{mathematically} admissible regions in the phase space, a trajectory that crosses into the region $0\leq r<1/2$ does not correspond to a \emph{physically viable} cosmological solution. The trajectory corresponding to a physically viable cosmological solution must always be in the regions $r<0$ or $r\geq 1/2$

It can be noticed that the function $\Lambda_c(r)$ is symmetric about $r=\frac{1}{2}$, which corresponds to the boundary value $\Omega=0$ (Eq.\eqref{condr2}) that we will show constitutes an invariant submanifold of the dynamics, and that $\frac{d\,\Lambda_c}{d\,r}\bigg\vert_{r\to1/2}=0$. $r=r(Q)$ is a dynamical quantity during the cosmic evolution, and so is $\Lambda_c(r)=\Lambda_c(r(Q))$. If, during the course of the cosmic evolution, $r(Q)\to1/2$, then $\Lambda_c\to0$ asymptotically.

In summary, at $\Lambda=\Lambda_{c}$ the system undergoes a saddle-node bifurcation,
where the two potential solutions coalesce into a single merged branch before disappearing (or appearing) entirely, depending on which side of the critical value the physical system resides.

\begin{figure}[H]
\centering 
\includegraphics[width=8cm,height=6cm]{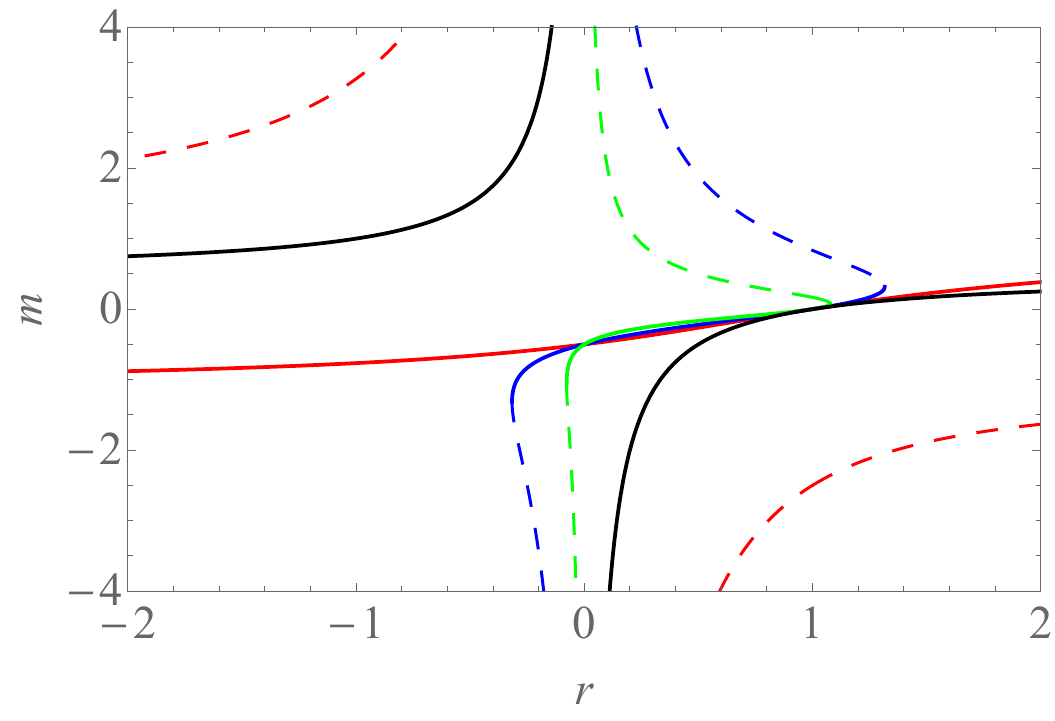}
	\caption{Plot of $m$ as function of $r$ from Eq.(\ref{mvsrG1}) for $\alpha=1, \beta=1$, where red color curve corresponds to $\Lambda=0.1$, blue colour for $\Lambda=0.2$, green colour for $\Lambda=0.5$ and black colour for $\Lambda=0$; the numerical values of $\beta$ and of $\Lambda$ are expressed in units of $H_0$ and $H_0^2$, respectively. The solid/dashed curves correspond to  positive/negative branches of $m$, respectively. We note that when $\Lambda \neq 0$ the function $m=m(r)$ is not anymore single-valued and the emergence of a continuous behavior between the two branches when increasing the value of the cosmological constant term $\Lambda$.   } \label{fig:m_r_plot_C1}
\end{figure}

We note that the two branches are generated by a non-zero value of the cosmological constant term $\Lambda$. We numerically inspect the behavior of the function $m=m(r)$ in Fig. \ref{fig:m_r_plot_C1} by identifying the following two noteworthy aspects with respect to the presence of a cosmological constant term $\Lambda$: {\it (i)} when $\Lambda \neq 0$, the function $m=m(r)$ is not anymore single-valued; {\it (ii)} below a certain numerical threshold for $\Lambda$, the two branches split realizing a second-order transition. Hence, the parameter $\Lambda$ supports (at least) these two types of bifurcations.

\subsection{The $m_i$-hierarchy approach}

To avoid the complexities of the double branch, we resort to the so-called $m_i$-hierarchy approach that was first introduced in \cite{Dutta:2025fqw} (In particular, see \cite[Sect.7]{Dutta:2025fqw}). Again, for a detailed exposition, we refer the readers to the Ref.\cite{Dutta:2025fqw}, while we just import the relevant equations here.

Defining the parameters
\beq\label{hierarchy}
m_i=Q \frac{d^{i+1} f/dQ^{i+1}}{d^i f/dQ^i}\,, \qquad m_1 \equiv m \,,
\eeq
it can be shown that for the first connection branch, they follow the recursion relation
(\cite[Eq.(121)]{Dutta:2025fqw}):
\beq\label{recursion}
m_i'=-3\Omega m_i\left(\frac{1-m_i+m_{i+1}}{1+2m_1}\right)\,.
\eeq

For the reconstructed $\Lambda$CDM-mimicking theory (\ref{flcdmG1}), the $m_i$-hierarchy stops at second order because $m_2=-3/2=const.$. Thus, the dynamical system is constituted by 
\begin{subequations}
	\label{closingtheoryG1}
	\begin{eqnarray}
		\Omega' &=& 3\Omega \left[\left(\frac{1 + m}{1 + 2m}\right)\Omega - 1\right]\,,
		\label{omp-mi}
		\\
		m' &=& \frac{3}{2} m \Omega\,. 
		\label{m1p-mi}
	\end{eqnarray}
\end{subequations}

Equations~(\ref{closingtheoryG1}) describe the coupled evolution of the effective matter density parameter and  the auxiliary function $m(Q)$ for the $\Gamma_1$ branch.  The first equation governs the energy exchange between the  matter and the geometrical sectors, while the second shows that 
the rate of evolution of $m$---which quantifies the deviation  from GR---is proportional to the matter content.  As the Universe approaches the vacuum regime ($\Omega\to0$), Eq.~(\ref{m1p-mi}) implies $m'\to0$, hence the geometry freezes to a constant $Q_0=-6H_0^2$, corresponding to a constant Hubble rate.  The system therefore asymptotes to a de~Sitter fixed point, consistent with the late-time $\Lambda$CDM behavior.

\begin{figure}[H]
	\centering
	\subfigure[]{%
		\includegraphics[width=8cm,height=6cm]{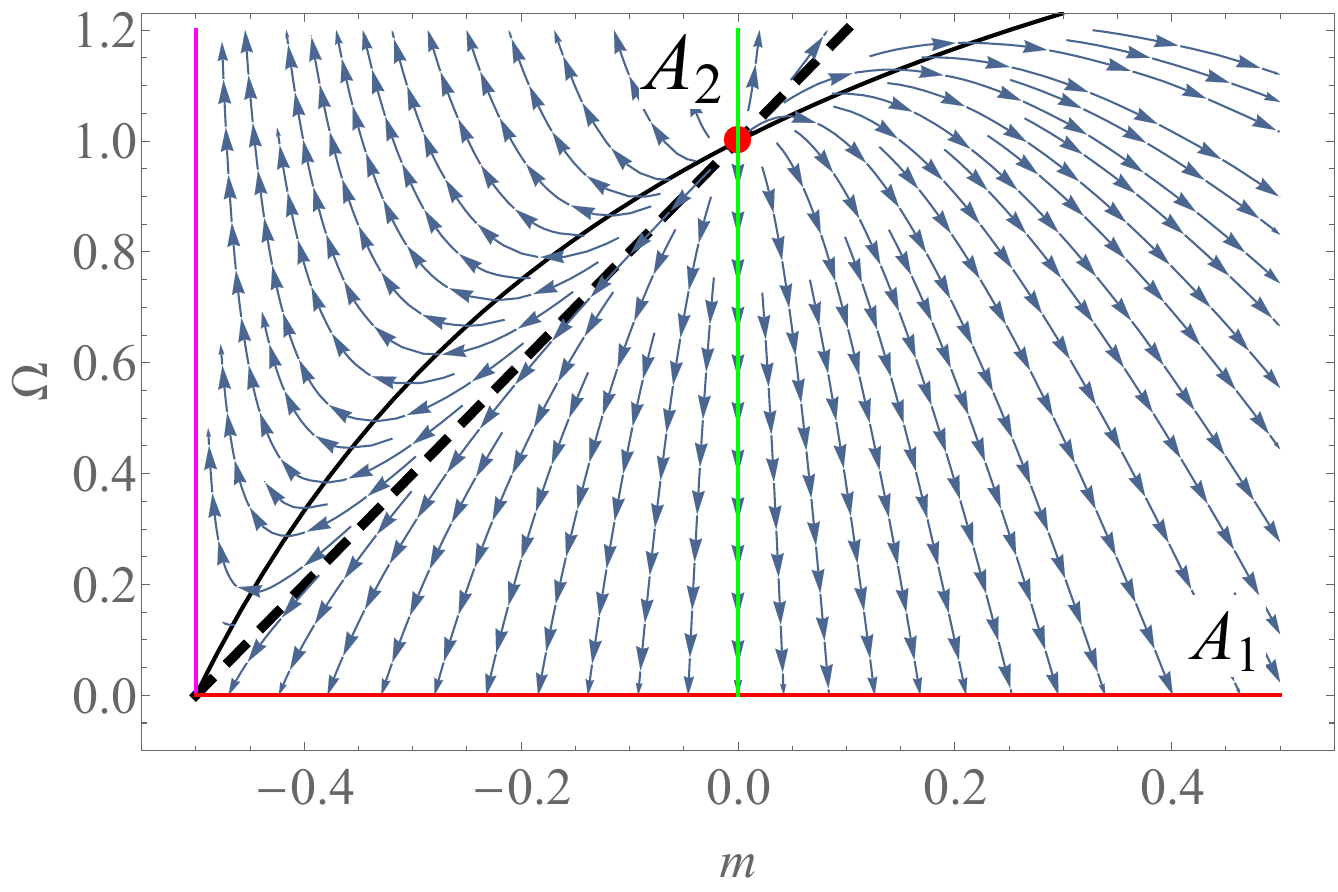}\label{fig:phase_theo_G1a}}
	\quad
	\subfigure[]{%
		\includegraphics[width=8cm,height=6cm]{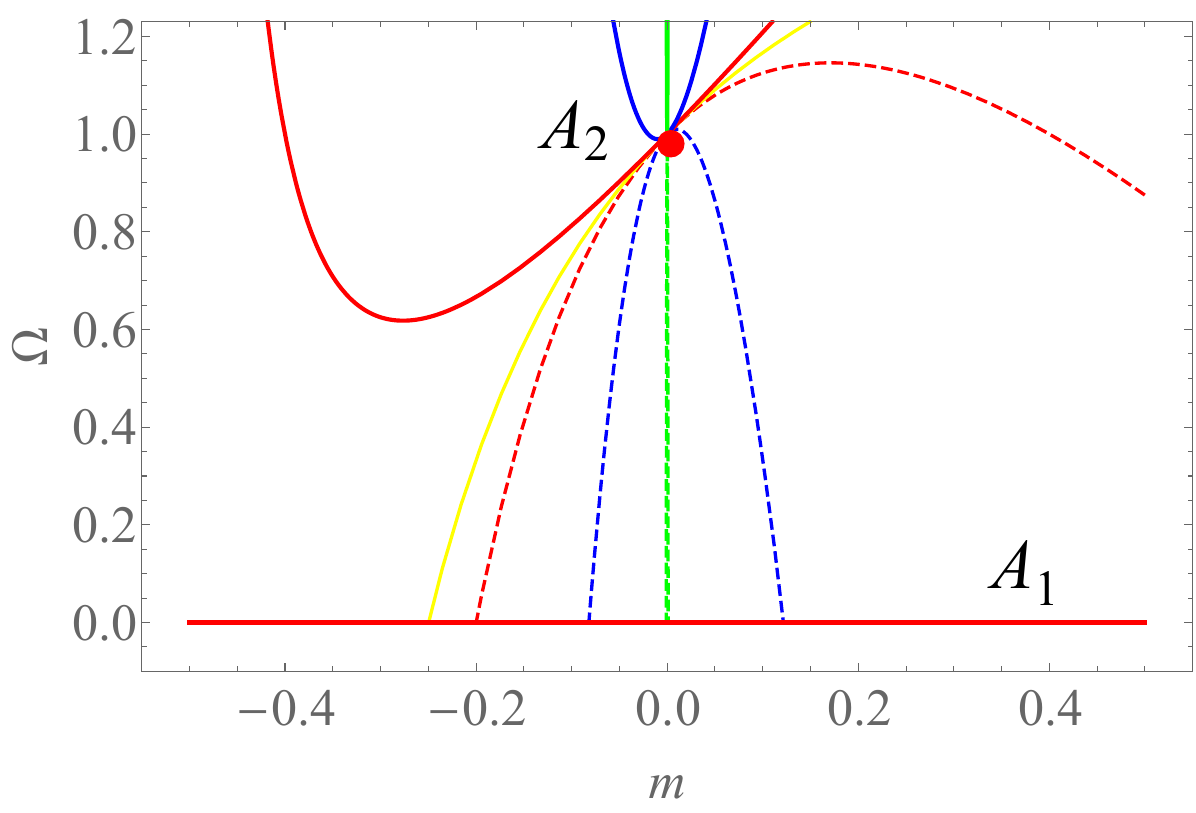}\label{fig:phase_theo_G1_soln}}
	\caption{In panel (a), we display the phase plot of the dynamical system (\ref{closingtheoryG1}). The horizontal nullcline is shown by the solid black line, the separatrix $\Omega=2m+1$ is shown by the dashed black line, and the invariant submanifolds $m=0,\,m=-1/2,\,\Omega=0$ are shown by the green, magenta, and red lines, respectively. The separatrix separates between two qualitative asymptotic behaviours of diminishing $\Omega$ and diverging $\Omega$. The two equilibria $A_1$ and $A_2$ correspond to the future asymptotic de-Sitter epoch and the past matter-dominated epoch, with the $\Lambda$CDM-mimicking trajectories connecting them. In panel (b), we plot the evolution of the effective matter abundance versus the auxiliary variable $m$, building upon our analytical solution (\ref{Omega_in_m}); we choose initial conditions such that $\mathcal{C}=10^6$ (solid green curve), $\mathcal{C}=-10^6$ (dashed green curve),  $\mathcal{C}=100$ (solid blue curve),  $\mathcal{C}=-100$ (dashed blue curve),  $\mathcal{C}=5$ (solid red  curve),  $\mathcal{C}=-5$ (dashed  red curve),  $\mathcal{C}=0$ (solid yellow curve). While all these curves cross the equilibrium epoch $A_2$, reaching $A_1$ depends on the initial conditions: this feature appears consistently across the two plots.}
		\label{fig:phase_theo_G1}
\end{figure}

It should be appreciated that the hierarchy approach, while preventing the difficulty of handling the two branches of the auxiliary variable, makes the dynamical system 2-dimensional. Moreover, while in the 1-dimensional dynamical system \eqref{DS_CT_G1} the model parameters $\alpha,\beta,\Lambda$ of the reconstructed function (\ref{flcdmG1}) appear explicitly through the expression $m(\Omega)$, they do not appear in the 2-dimensional dynamical system \eqref{closingtheoryG1} at all. The roles of the numerical values of the free parameters on the qualitative features of the dynamics are washed away. In a sense, the 2-dimensional dynamical system \eqref{closingtheoryG1} that we have reached via the hierarchy approach reveals the generic dynamical features of the reconstructed theory \eqref{flcdmG1}, which is \emph{independent} of the model parameters. This revealing of the generic \emph{parameter-independent} feature is a distinctive advantage of the hierarchy approach over the standard approach, giving rise to the parameter-dependent 1-dimensional system \eqref{DS_CT_G1}. The resulting generic phase portrait is presented in Fig. (\ref{fig:phase_theo_G1a}).

Let us mention at this point that the kind of \emph{branching} that we observed in Sec.\ref{ss2a} is, in a way, present, but hidden, in the hierarchy approach. Recall that the Friedmann constraint delivers $r=\frac{1}{2-\Omega}$. If one wants to obtain $Q=Q(\Omega)$ by substituting $r(Q)$ from \eqref{rvsQG1} into the Friedmann equation, one runs once again into a double-valued function $Q(\Omega)$. However, this double-valued-ness does not actually affect the generic features of the cosmological dynamics itself, as no such branching appears in the system \eqref{closingtheoryG1}.

We also note that $m$ is a monotonically increasing/decreasing function and $m=0$ is a globally repelling invariant submanifold. In the $\Lambda$CDM-mimicking scenario, $Q=-6H^2\neq0$, so that $m\to0$ implies the GR limit. The fact that the invariant submanifold $m=0$ is always repelling implies that GR acts as a \emph{cosmological past attractor} for cosmological dynamics under this theory. 

The Eq.\eqref{omp-mi}, and hence the dynamical system \eqref{closingtheoryG1} as a whole, appears to be singular at $m=-1/2$. However, one can see from Eq.\eqref{mvsrG1} that the only possibility for $m=-1/2$ is $\alpha=0$ (since $Q=-6H^2\neq0$ in a late-time expanding universe). Since we have started with the very assumption that $\alpha,\beta\neq0$ for the theory under consideration \eqref{flcdmG1}, $m=-1/2$ is never reached. And hence, the system \eqref{closingtheoryG1} is always regular in its domain of application. In fact, $m=-1/2$ appears as an 1-dimensional invariant submanifold in the phase space $m-\Omega$; see Fig.\ref{fig:phase_theo_G1}.

We notice that the vacuum solutions with $\Omega=0$ form a line of attractors in the phase space. Moreover, on this line, we can compute $Q=-2\Lambda/\alpha\Rightarrow H=\sqrt{\Lambda/(3\alpha)}$ by combining (\ref{rvsQG1}) with (\ref{condr2}). The existence of the vacuum solutions, therefore, necessitates
\beq\label{lambda_bound}
\frac{\Lambda}{\alpha}>0\,.
\eeq 
We will re-obtain this restriction below when studying the invariant submanifold $m=0$.

Note that, although in this paper we are denoting the model parameter $\Lambda$ as the cosmological constant, the actual \emph{physical} role of the cosmological constant is played by the ratio $(\Lambda/\alpha)$. This can be seen the fact that the abundance parameter that asymptotically becomes one in the future asymptotic epoch is actually the quantity $\frac{(\Lambda/\alpha)}{3H^2}$.

In the region of the phase plane $m>-1/2$, the generalized hyperbola $\frac{1+m}{1+2m}\Omega=1$ can be identified as the horizontal nullcline\footnote{A horizontal nullcline is where the vertical component of the phase flow vanishes, so that the flow becomes completely horizontal.}, and the line $\Omega=2m+1$ can be identified as a separatrix. The separatrix $\Omega=2m+1$ is an important discriminator of the qualitative behaviours of the phase trajectories; see Fig.\ref{fig:phase_theo_G1}. The basin of attraction for the line of attractors $\Omega=0$ lies in the region $\Omega<2m+1$; any initial condition chosen below this separatrix leads to a trajectory that ultimately crashes onto the line $\Omega=0$. On the other hand, any initial condition chosen above this separatrix leads to a trajectory with diverging $\Omega$. Astrophysically meaningful initial data for interpolating between matter-dominated and vacuum-dominated epochs should definitely lie below the separatrix $\Omega=2m+1$.

\subsection{The fixed points}

Equations~(\ref{closingtheoryG1}) yield two equilibria: a dark-energy dominated line of fixed points $A_1(\Omega,\,m)=(0,\,const.)$, and a dark matter dominated isolated fixed point $A_2 (\Omega,\,m)=(1,\,0)$, respectively\footnote{Note that, $Q(A_2)=-(4\Lambda/\beta)^2<0$,  by equating (\ref{rvsQG1}) with (\ref{condr2}).}. Thus, with $n=1/2$, we have obtained one equilibrium fewer than the case with the general considered in \cite[Table 8]{Dutta:2025fqw}, along with a phase space bifurcation presented by a nonzero cosmological constant. In proximity of the line of fixed points $A_1(\Omega,\,m)=(0,\,const.)$, the linear term becomes subdominant in the theory (\ref{flcdmG1}), while in proximity of the isolated fixed point $A_2 (\Omega,\,m)=(1,\,0)$, the square root term becomes subdominant. 

Let us now deepen the understanding of the two points by commenting on the following
\begin{enumerate}
\item Physical meaning of the square root term in the reconstructed theory. The $\beta\sqrt{-Q}$ term in the $f(Q)$ theory actually does not have any dynamical contribution on its own for a homogeneous and isotropic background. At the background FLRW level, this part of the theory contributes to a boundary term, which can be explicitly shown using the corresponding point-like minisuperspace Lagrangian; see Appendix \ref{appA}. Therefore, at the background level, the theory (\ref{flcdmG1}) will produce the same FLRW dynamics as $f(Q)=-2\Lambda+\alpha Q$, i.e., STEGR. This conclusion, however, does not hold for other backgrounds or even at the perturbed FLRW level; the latter case provides an opportunity to break the degeneracy \cite{Albuquerque:2022eac}.
\item Physical meaning of $m=0$: $m\equiv\frac{Q f_{QQ}}{f_Q}=0$. Since $Q=-6H^2$ and, in a late time cosmology, one typically does not encounter a situation with $H=0$\footnote{We keep a turnaround situation $H=0,\,\dot{H}<0$, that may appear in very specific cosmological models, out of consideration. Such a scenario can not be tracked by the finite fixed point analysis in terms of Hubble normalized variables anyway.}, $m=0\Rightarrow f_{QQ}=0\Rightarrow f_Q=constant$, which is the GR limit of the theory. For the particular theory under consideration, one can compute $f_{QQ}=-\frac{\beta}{4(-Q)^{3/2}}=-\frac{\beta}{24\sqrt{6}H^3}$, so that this GR limit ($m\to0$) of the theory is achieved in the limit $H\to\infty$. Luckily, this is a runaway singularity because it takes an infinite amount of redshift, and consequently of cosmic time, for the Hubble function to diverge in a $\Lambda$CDM-mimicking cosmology (\ref{HLCDM}). On the other hand, by using Eq.(\ref{trading}) below, we see that the deceleration function at the GR limit is finite, reading as $q(z\to\infty)=\frac{3 \Omega(z\to\infty)}{2}-1$, and we have consistently $q(\Omega=1)=\frac{1}{2}$.

The way we have written the $f(Q)$ action here, the effective gravitational coupling in the vicinity of this fixed point appears in general to be different than unity, reading as
$$\lim_{z\to\infty}f_Q=\alpha \Rightarrow \lim_{z\to\infty}\kappa_{\rm eff}=(\alpha)^{-1}.$$ 
However, this only amounts to a constant rescaling of units, not affecting physical conclusions, because we have taken the matter field to be minimally coupled to geometry. However, $\alpha$ must be positive. Indeed, we have, in the vicinity of this point, 
\beq\label{alpha_bound}
1=\Omega(z\to\infty)=\lim_{z\to\infty}\frac{\Omega_{m0}(1+z)^3}{h^2(z)f_Q(z)}=\frac{3\Omega_{m0}}{2\alpha} \quad \Rightarrow \quad \alpha=\frac{3}{2}\Omega_{m0}>0\,,
\eeq
where we have used the fact that for a $\Lambda$CDM-like evolution $ \displaystyle \lim_{z\to\infty}\frac{2}{3}(1+q_0)(1+z)^3$. 
\end{enumerate}


\subsection{Analytical solution for the evolution of the  matter abundance}

Lastly, by taking the ratio side by side of the dynamical equations (\ref{closingtheoryG1}),  we can explicitly obtain the following closed-form solution 
\beq \label{Omega_in_m}
\Omega(m)=\frac{\mathcal{C}m^2+4m+1}{2m+1}\,, \qquad {\mathcal C} =\frac{(2m_0+1)\Omega_0-4m_0-1}{m_0^2}\,,
\eeq
which shows that the dynamics is effectively one-dimensional even though the hierarchy approach, that is, the intermediate step we have relied on, is constituted by two equations. 
Different phase orbits correspond to different choices of the arbitrary integration constant ${\mathcal C}$, e.g. of initial conditions here denoted with a subscript 0. We have a transition to a straight line orbit $\Omega(m)=2m+1$ should the initial conditions be such that $\Omega_0=2m_0+1$, (e.g., $ {\mathcal C}=4$); this latter solution is the separatrix. We also remark that this analytical solution holds outside the equilibrium points at which our procedure of taking the ratio side by side of the dynamical equations breaks down. We certify numerically in Fig. \ref{fig:phase_theo_G1_soln} the correctness of our results for the model evolution by comparing the dynamics in the phase plane with our obtained solution for the matter abundance.

Solution (\ref{Omega_in_m}) is consistently recovered from the Friedmann constraint (note that $x_1=x_1(Q(m)))$ via Eqs.(\ref{dynvar_gamma1}) and (\ref{mvsrG1}) if we identify 
\beq
\label{CC1}
{\mathcal C}=4\left(1 -8\frac{\alpha\Lambda}{\beta^2}\right) \,.
\eeq
This identification shows that different trajectories in the phase plane originating from different initial conditions can be re-interpreted as corresponding to different numerical values of the particular model parameter combination $\frac{\alpha\Lambda}{\beta^2}$. The role of this particular combination of the model parameters could not be discerned at first sight from the dynamical system formulation. The particular case of the separatrix solution $\Omega=2m+1$, which arises for $\mathcal{C}=4$ and constitutes a transition to a one-to-one relationship between the two dynamical variables, is achieved in the limit $\frac{\alpha\Lambda}{\beta^2}\to0$, and it corresponds to a motion with initial conditions $\Omega_0=1+2m_0$. 

Furthermore, using (\ref{mvsrG1}) with (\ref{Q-1}) and (\ref{HLCDM}), we can eventually obtain from the analytical solution \eqref{Omega_in_m} the following redshift evolution of the effective matter abundance:
\beq
\label{OOR1}
\Omega(z)=  \frac{4\left[\frac{\Lambda}{H_0^2} - 3\,\alpha\,h^2(z)\right]}{\sqrt{6}\left[\frac{\beta}{H_0} - 2\sqrt{6}\,\alpha\,h(z)\right]h(z)}\,, \qquad\qquad h^2(z)=\frac{2}{3}(1+q_0)(1+z)^3 + \frac{1}{3}(1-2q_0)\,.
\eeq
The physical interpretation of our result is therefore the following: we could obtain a cosmological model with the same cosmographic properties of the GR $\Lambda$CDM model, but with a completely different evolution of its matter abundance. We will put this finding in the context of the cosmological tensions revealed by recent astrophysical analyses in Sect.\ref{ss4}.

\section{Cosmography-Driven Closure}
\label{ss2b}

The cosmographic approach to cosmology builds upon a tower of algebraic combinations of the Hubble rate and its derivatives. The first two parameters, and the ones that will be required for our analysis, are the deceleration parameter $q$ and the jerk parameter $j$ \cite{Visser:2003vq,Dunajski:2008tg,Capozziello:2019cav}: 
\begin{subequations}
	\begin{eqnarray}
		q &\equiv& -\frac{1}{aH^2}\frac{d^2 a}{dt^2} = -1-\frac{\dot H}{H^2} \,,
		\\
		\label{jerkfunct}
		j &\equiv& \frac{1}{aH^3}\frac{d^3 a}{dt^3} = \frac{\ddot H}{H^3} - 3q - 2 \,,
	\end{eqnarray}
\end{subequations}
which are interrelated by the relation
\begin{equation}
    q' = 2q^2+q-j\,.
\end{equation}
The GR $\Lambda$CDM model is kinematically specified by the cosmographic condition $j(z)=1$, which, it is suggested, is consistent with datasets from cosmic chronometers, supernovae, and baryon acoustic oscillations \cite{Mukherjee:2016shl,Bernal:2016gxb,Mukherjee:2020ytg,Jiang:2024xnu,Gao:2025ozb}. 

Integrating back from the cosmographic condition $j(z)=1$, it is possible to analytically obtain the evolution of the Hubble parameter \cite{Zhai:2013fxa,Chakraborty:2022evc,Chakraborty:2025qlv} 
\beq
\label{HLCDM}
\frac{H^2(z)}{H_0} = \frac{2}{3}(1+q_0)(1+z)^3 + \frac{1}{3}(1-2q_0)\,,
\eeq
and the deceleration parameter
\begin{subequations}\label{qlcdm}
\begin{align}
    q(z) &= \frac{c_1 (1+z)^3 -2(1-c_1)}{2[c_1(1+z)^3 +1-c_1]},~~~c_1=\frac{2}{3}(1+q_0)\,,
    \\
    q(N) &= \frac{1+C_1 e^{3N}}{2-C_1 e^{3N}},~~~C_1=2-\frac{2}{c_1}=-\frac{1-2q_0}{1+q_0}\,.
\end{align}
\end{subequations}

We also recall that the applicability of (\ref{HLCDM}) lies in the range \cite[Sect.II]{Chakraborty:2025qlv}:
\beq
\label{rangeq}
-1 \leq q \leq \frac{1}{2}\,.
\eeq

When we talk about a $\Lambda$CDM-mimicking theory, like in this paper, we imply the possibility of achieving the cosmographic condition $j(z)=1$ within the framework of that theory (see, e.g., \cite{Chakraborty:2021jku,Chakraborty:2022evc,Chakraborty:2025qlv}).

\begin{figure}[H]
\centering
	\subfigure[]{%
		\includegraphics[width=8cm,height=6cm]{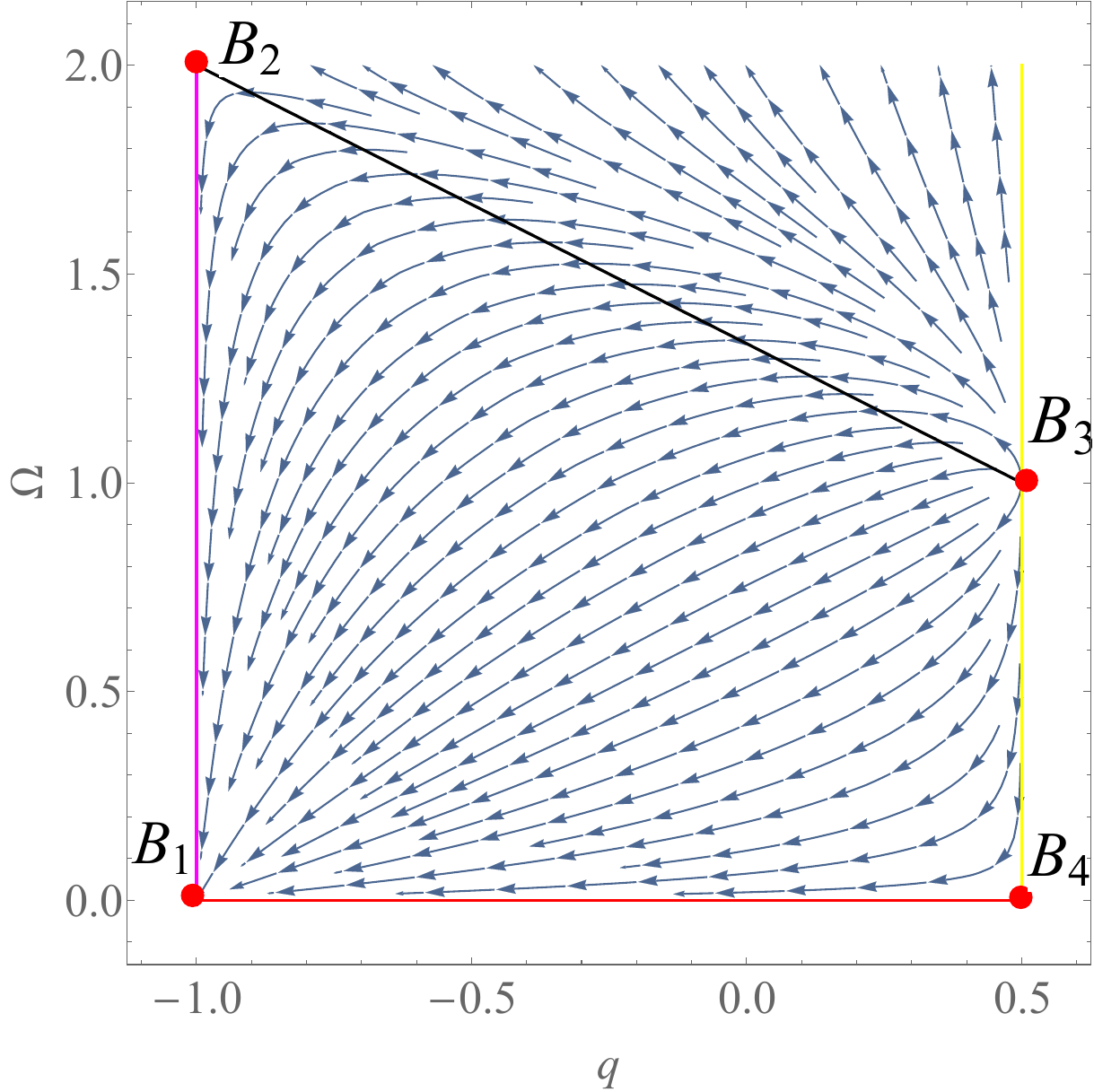}\label{fig:phase_cosmo_G1a}}
	\quad
	\subfigure[]{%
		\includegraphics[width=8cm,height=6cm]{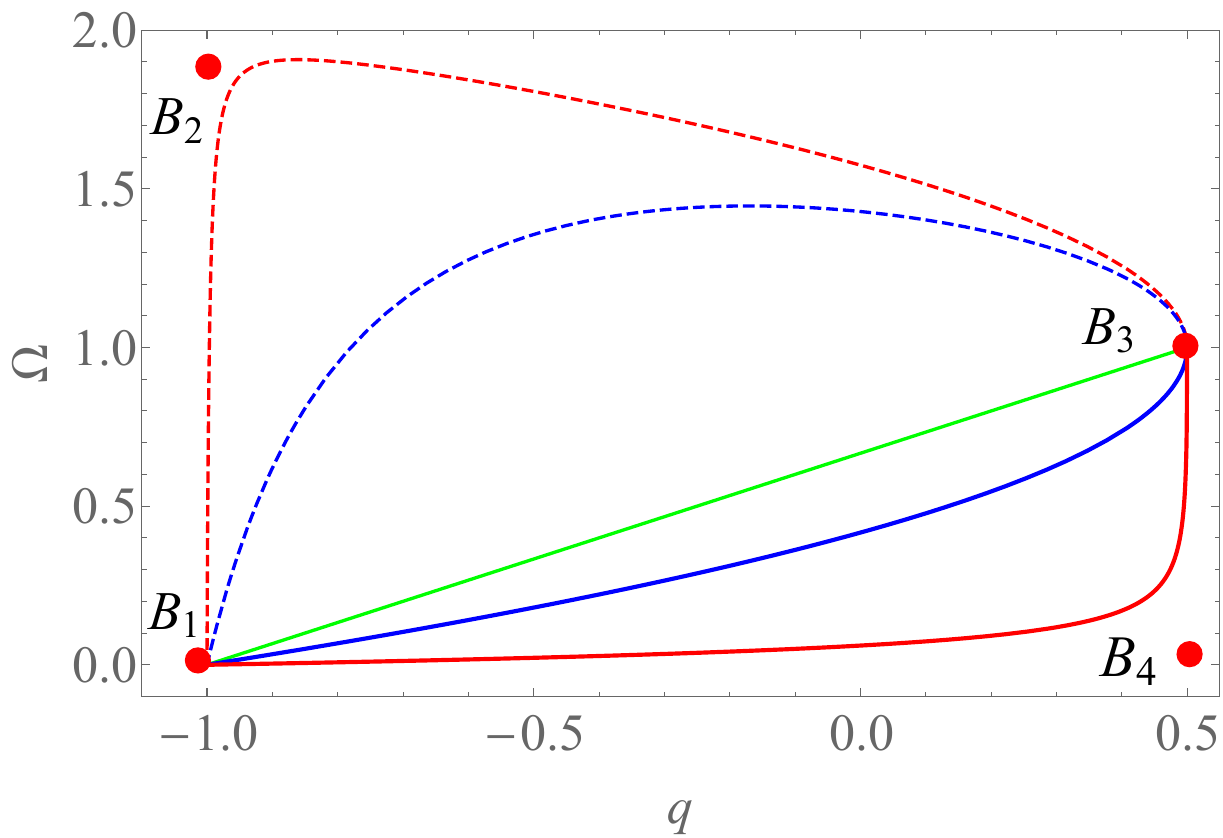}\label{fig:phase_cosmo_G1_soln}}
	\caption{In panel (a), we present the phase dynamics of the system (\ref{evcosmoG1}) whose equilibria denoted by  red bullets are listed in table \ref{tab:G1_cosmo_pts}; the purple, yellow and red lines identify three invariant submanifolds which bound the physically relevant portion of the phase plane via the constraints (\ref{rangeq}) and $\Omega \geq 0$. The heteroclinic orbit from $B_3$ to $B_2$ connects the two boundaries $q=1/2$ and $q=-1$ in the phase space. The black line given by $\Omega =\frac{2(2-q)}{3}$, constituting a heteroclinic orbit between  $B_3$ and $B_2$, separates the regions in which $\Omega$ increases (above it) and decreases (below it). We corroborate these identified qualitative behaviors by plotting in panel (b) our analytical solution (\ref{Omegavsq})  with the initial conditions  $\tilde{\mathcal{C}}=-1.73$ (red dashed curve), $\tilde{\mathcal{C}}=-1.6$ (blue dashed curve), $\tilde{\mathcal{C}}=0$ (green curve), $\tilde{\mathcal{C}}=1.8$ (blue solid curve), $\tilde{\mathcal{C}}=30$ (red solid curve). We remark that the trajectories of the mimicking $\Lambda$CDM cosmological model are those connecting $B_3$ to $B_1$.} \label{fig:phase_cosmo_G1}
\end{figure}

We can trade the auxiliary variable $m$ in terms of the cosmographic deceleration parameter via the following relation:
\beq
\label{trading}
m= \frac{1}{2}\left[ \frac{3\Omega}{2(1+q)}  -1 \right]\,.
\eeq
The above equation can also be directly obtained from the field equations (see, e.g. \cite[Eq.(26)]{Dutta:2025fqw}). This operation of trading shall preserve the dimensionality of the phase space. Then, the dynamical system is constituted by:
\begin{subequations}
	\label{evcosmoG1}
	\begin{eqnarray}
		\label{evcosmoG1a}
		\Omega' &=& \frac{3}{2}\Omega\left[\Omega-\frac{2}{3}(2-q)\right]\,,\\
		q' &=& 2 q^2+q-1\,,
	\end{eqnarray}
\end{subequations}
where we have enforced the condition $j=1$ in the evolution of the deceleration function. We note that, unlike in the \lq\lq closing with the theory" formulation (\ref{closingtheoryG1}), in the ``closing with cosmography'' approach, one of the equations (the $q'$-equation) decouples. Hence, after having enforced a specific cosmological evolution via a certain cosmographic condition such as $j=j(q)$, we are left effectively with only one dynamical evolution equation; that for the effective matter abundance parameter $\Omega$. We recall that the $\Lambda$CDM-mimicking ($j=1$) deceleration function is monotonically decreasing, while $\Omega$ is increasing/decreasing for $\Omega \gtrless \frac{2}{3}(2-q)$, respectively; we will see in Fig. \ref{fig:phase_cosmo_G1a} that the equality applies to a heteroclinic orbit connecting two equilibria; moreover, by using Eq.(\ref{trading}), this corresponds to what we dubbed \lq\lq generalized hyperbola" in the context of closing the dynamical system with the theory.

It appears that the only difference between the systems \eqref{closingtheoryG1} and \eqref{evcosmoG1} is that the phase space dimension $m$ is replaced by $q$. Thus, whereas the closing with theory approach was suited for showing how the deviation of the theory from GR evolves with time, the closing with cosmography approach enlightens the transition between cosmological epochs characterized by different $q$-values.

The above system \eqref{evcosmoG1} has four fixed points that we list in Table \ref{tab:G1_cosmo_pts}.
	\begin{table}[H]
		\centering
	\renewcommand{\arraystretch}{1.6}
		\begin{tabular}{|c|c|c|c|c|}
			\hline 
			Fixed point  & $(\Omega, q)$ & Stability  &  Cosmological solution  \\ 
			\hline 
			$B_1$ & $(0,-1)$ & Attractor & $a(t)=a_0 e^{H_0 t}$ \\
			\hline			
			$B_2$ & $(2,-1)$ & Saddle & $a(t)=a_0 e^{H_0 t}$ \\
			\hline	
			$B_3 $ & $\left(1 ,\frac{1}{2}\right)$ & Repeller  & $a=a_0 t^{\frac{2}{3}}$\\ 		
			\hline 
			$B_4$ & $\left(0,\frac{1}{2}\right)$ & Saddle & $a=a_0 t^{\frac{2}{3}}$\\ 
			\hline 		
		\end{tabular}
		\caption{In this table, we list the fixed points for the system \eqref{evcosmoG1} together with their stability and the cosmology they support. $B_{2,4}$ could not be identified in Sect. \ref{ss2a} because they correspond to the pathological cases $m=\infty$ and $m=-1/2$ for the auxiliary variable; nevertheless, they should be excluded when taking into account boundary conditions for the dynamics.}
		\label{tab:G1_cosmo_pts}
	\end{table}
Fixed points $B_1$ and $B_2$ correspond to an accelerated de-Sitter solution, out of which only the point $B_1$ corresponds to a vacuum solution. The fixed point $B_3$ behaves as a matter-dominated repeller solution. The point $B_4$ corresponds to a certain vacuum solution whose scale factor evolves the same way as that of a GR matter-dominated universe. The fixed points $B_2$ and $B_4$ could not be identified via the \lq\lq closing with theory" approach of Sec.\ref{ss2a}, because they correspond to $m\to\infty$ and $m=-1/2$ respectively; both these cases were kept out of consideration in Sec.\ref{ss2a}. The fixed points $B_1$ and $B_3$ are clearly equivalent to $A_1$ and $A_2$ obtained in Sec.\ref{ss2a}. 

Although it seems like the \lq\lq closing with cosmography" approach can identify more fixed points, namely $B_{2,4}$, in the solution space of the underlying theory, they are not of much physical interest. Only the trajectories directly connecting $B_3$ to $B_1$ are cosmologically relevant. The phase portrait of the system \eqref{evcosmoG1} is shown in Fig.\ref{fig:phase_cosmo_G1a}. 

\subsection{Analytical solution of the dynamical system}\label{subsec:analytic}

The orbits in the phase plane read as\footnote{The following is a Bernoulli equation $y'(x)+P(x)y+Q(x)y^n=0$ with $n=2$.}:
\begin{eqnarray}
\label{Omegavsq}
\frac{d \Omega}{dq}=\frac{\Omega'}{q'}=\frac{[3\Omega+2(q-2)]\Omega}{2(2 q^2+q-1)} \qquad \,\, \,\,\Rightarrow \qquad \,\,\,\, \Omega(q)=\frac{2(1+q)}{3 + \tilde{\mathcal C}\sqrt{1-2q}}\,, \quad \tilde{\mathcal C}=\frac{2(1+q_0)-3\Omega_0}{\sqrt{1-2q_0}\Omega_0}.
\end{eqnarray}
Since the deceleration function of $\Lambda$CDM-mimiking cosmology, e.g., when $j=1$, is a known function of $N$ or $z$ (see Eq.\eqref{qlcdm}), Eq.\eqref{Omegavsq} can also be expressed in two other equivalent forms:
\begin{subequations}
\begin{eqnarray}
\Omega(N)& =& \frac{2}{2 - C_1 e^{3N} +2C_2 e^{3N/2}\sqrt{\left(2-C_1 e^{3N}\right)}}, \qquad C_1=-\frac{1-2q_0}{1+q_0}\,\,,\, C_2 = \frac{2(q_0 + 1) - 3\Omega_0}{2\Omega_0 \sqrt{3(q_0 + 1)}}\,, \label{Om_sol_G1_N}\\
\Omega(z) &=& \frac{2(1+z)^3}{2(1+z)^3 - C_1 + 2C_2 \sqrt{2(1+z)^3 - C_1}} \,,\qquad C_1=-\frac{1-2q_0}{1+q_0}\,\,,\, C_2 = \frac{2(q_0 + 1) - 3\Omega_0}{2\Omega_0 \sqrt{3(q_0 + 1)}}\,. \label{Om_sol_G1_z}
\end{eqnarray}
\end{subequations}

Let us now inspect the constituting blocks on which $\Omega$ is based. The redshift evolutions of the Hubble function and of the matter density are as in the GR $\Lambda$CDM model. The former is one of the core assumptions of our present paper; the latter follows from the minimal coupling between matter and geometry, which preserves the characteristics of the general relativistic geodesic motion. Therefore, the difference in the matter abundance parameters is fully contained in the effective coupling. At the present day, we can interpret it as playing a dressing effect for the matter density. Considering the value of $H_0$ as given (for example, obtained in a model-independent manner from cosmic chronometers data), ameliorating the Hubble tension requires decreasing the present-day energy density of pressureless matter \cite{Bernal:2016gxb}. In our framework, this can be \lq\lq effectively" achieved by a smaller than one value of the gravitational constant. While the gravitational constant constitutes an overall multiplicative factor in the background field equations with no measurable physical consequences, this is no longer the case at the perturbative level \cite{Li:2025msm}. In the specific context of $f(Q)$ gravity, a smaller value of the gravitational coupling is consistently required also for a realistic clustering of astrophysical structures \cite{Boiza:2025xpn}.

We remark that our analytical solution can account for the different trajectories between the equilibria depending on initial conditions, but not for the equilibria themselves for which $\Omega'=0=q'$, making Eq.\eqref{Omegavsq} ill-defined. We depict our obtained solution in Fig.\ref{fig:phase_cosmo_G1_soln}, certifying that it correctly interpolates between the matter and vacuum dominated epochs listed in Table \ref{tab:G1_cosmo_pts}. 

Lastly, as a consistency check, we note that by trading the deceleration parameter in terms of the auxiliary variable $m$ via (\ref{trading}), we get back

\beq\label{one-to-one}
\Omega(m)=\frac{\left(4-\frac{12}{\tilde{\mathcal{C}}^2}\right) m^2+4m+1}{2m+1}\,,
\eeq
which is the same as \eqref{Omega_in_m} on the identification\footnote{This shows that the transition to a straight line for the solution $\Omega=\Omega(m)$ in \eqref{Omega_in_m} would require $\tilde{\mathcal{C}} \to \infty$, which can happen only at boundary values $q_0=1/2$ or $\Omega_0=0$. On the other hand, the condition $\tilde{\mathcal{C}}=0$, which would provide a linear relation $\Omega=\Omega(q)$ (from which bounds on the matter abundance would be straightforwardly set from those on the deceleration function), corresponds to  $m_0=0$, which in turn provides the invariant submanifold previously analyzed.} $\mathcal{C}=4-\frac{12}{\tilde{\mathcal{C}}^2}$. 

This relation establishes a correspondence between the integration constants of the two closure schemes. Therefore, the analytical solutions obtained from the theory-driven and cosmography-driven formulations describe precisely the same family of phase-space trajectories, despite being expressed in different variables. This confirms that the cosmographic closure does not introduce additional solutions but instead provides an alternative parametrization of the underlying dynamical system.

It is to be noted that the heteroclinic trajectory connecting $B_3$ and $B_1$ via $B_2$ and the one connecting $B_3$ and $B_1$ via $B_4$ enclose the basin of attraction for the late-time de-Sitter attractor $B_1$. All the physically viable $\Lambda$CDM-mimicking $f(Q)$ trajectories must reside within this basin. Physically, this means that, if one wants to set an initial condition at around the past attractor $B_3$ and obtain a physically viable $\Lambda$CDM-mimicking $f(Q)$ cosmological solution numerically, there is an upper limit on the initial condition of $\Omega$. 

It should be appreciated that dynamical system in the closing with cosmography approach, eq.\eqref{evcosmoG1}, allows us to analytically solve for all the mathematical quantities of interest. We have already expressed the solutions for $\Omega$ in eqs.\eqref{Omegavsq},\eqref{Om_sol_G1_N} and \eqref{Om_sol_G1_z}. Since, for $\Gamma_1$, $r=\frac{1}{2-\Omega}$ (Eq.\eqref{condr2}), this allows us to analytically solve for $r$:
\begin{subequations}
\begin{align}
r(N) &= \frac{2 - C_1 e^{3N} + 2C_2 e^{3N/2}\sqrt{2 - C_1 e^{3N}}}{2 \left( 1 - C_1 e^{3N} + 2C_2 e^{3N/2}\sqrt{2 - C_1 e^{3N}} \right)}, 
\\
r(z) &= \frac{2(1+z)^3 - C_1 + 2C_2(1+z)^{3/2}\sqrt{2(1+z)^3 - C_1}}{2\left((1+z)^3 - C_1 + 2C_2(1+z)^{3/2}\sqrt{2(1+z)^3 - C_1}\right)}, 
\\
r(q) &= \frac{3 + \tilde{\mathcal{C}}\sqrt{1-2q}}{2(2 - q + \tilde{\mathcal{C}}\sqrt{1-2q})}.\label{r(q)_G1}
\end{align}
\end{subequations}

Next, since the analytical solutions of $\Omega$ and $q$ are already known, the trading equation \eqref{trading} allows us to obtain analytical solutions for $m$:
\begin{subequations}\label{msoln_G1}
\begin{align}
m(N) &= \frac{-C_2 e^{3N/2}\sqrt{2 - C_1 e^{3N}}}{2 - C_1 e^{3N} + 2C_2 e^{3N/2}\sqrt{2 - C_1 e^{3N}}}, 
\\
m(z) &= \frac{-C_2(1+z)^{3/2}\sqrt{2(1+z)^3 - C_1}}{2(1+z)^3 - C_1 + 2C_2(1+z)^{3/2}\sqrt{2(1+z)^3 - C_1}}, 
\\
m(q) &= \frac{-\tilde{\mathcal{C}}\sqrt{1-2q}}{2(3 + \tilde{\mathcal{C}}\sqrt{1-2q})}.\label{m(q)_G1}
\end{align}
\end{subequations}

\begin{figure}[H]
    \centering
   \subfigure[]{%
		\includegraphics[width=8cm,height=6cm]{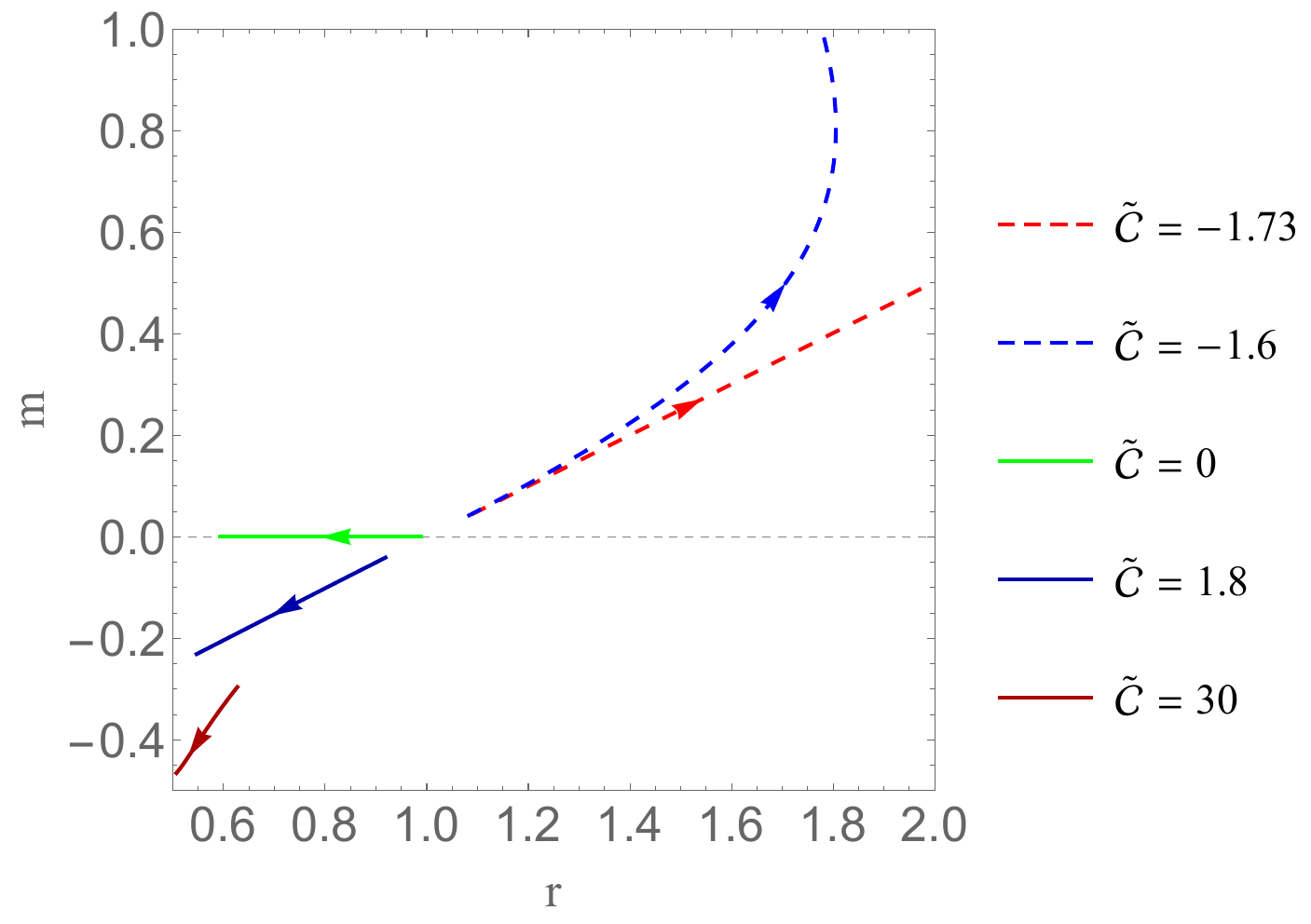}\label{fig:m_r_plotG1_analytical}}
	\quad
	\subfigure[]{%
		\includegraphics[width=8cm,height=6cm]{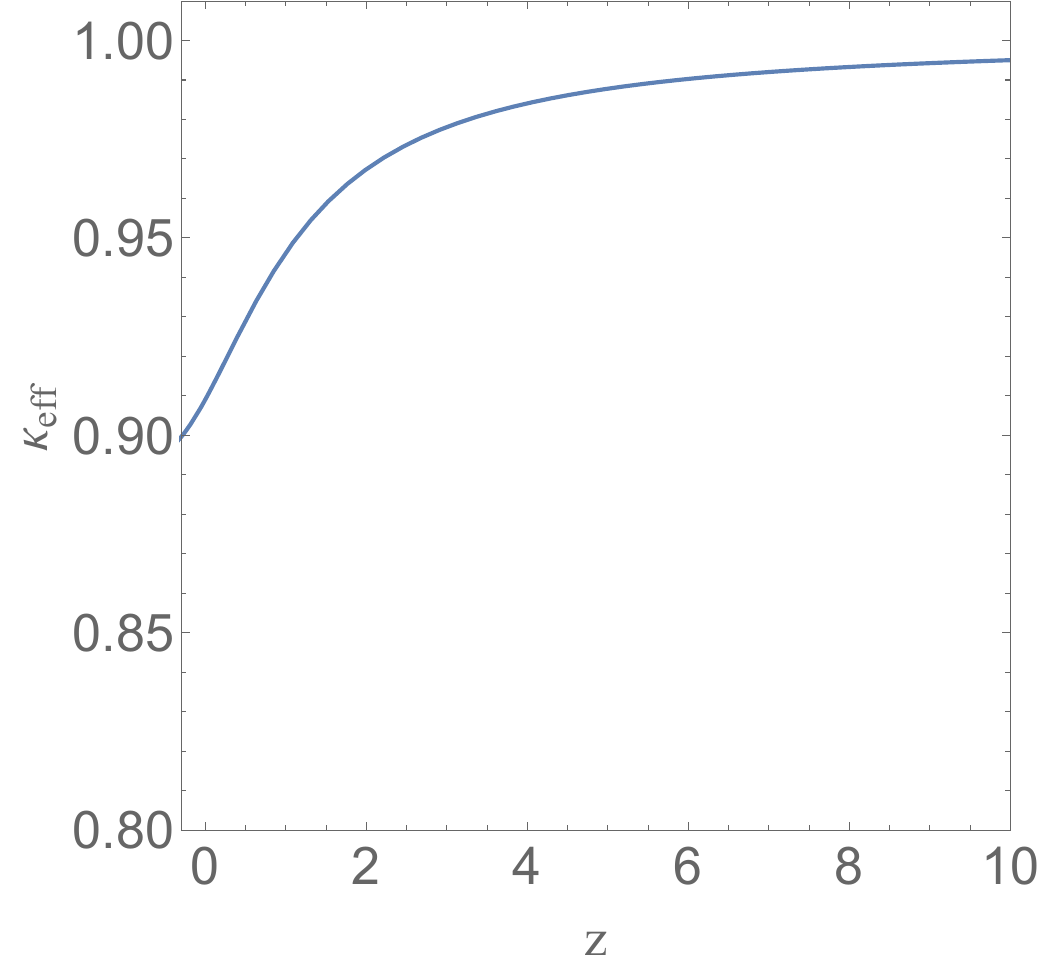}\label{fig:keff_c1}}
    \caption{Panel \ref{fig:m_r_plotG1_analytical} shows the plot of $m(q)$ as function of $r(q)$ for different values of $\tilde{\mathcal{C}}$. The values of $\tilde{\mathcal{C}}$ are the same as the ones considered in Fig. \ref{fig:phase_cosmo_G1}. The figure is obtained by a parametric plot $r(q)$ vs $m(q)$ within the range $q=0.49$ to today ($q=-0.55$), where the analytical solutions of $r(q)$ and $m(q)$ are taken from Eqs.\eqref{r(q)_G1} and \eqref{m(q)_G1} respectively. Panel \ref{fig:keff_c1} shows the evolution of the effective gravitational coupling $\kappa_{\rm eff}$ versus redshift for $q_0=-0.55$ and $\Omega_0=0.27$ ($\Omega_0=0.3$ makes $\kappa_{\rm eff}=1$ identically, i.e. the GR $\Lambda$CDM model). Both the panels in serves to show a stable GR regime in the high-redshift where $m(r)\to0$ and $\kappa_{\rm eff} \to 1$. As the universe transitions into the late-time accelerated expansion phase, the deviation from GR becomes increasingly pronounced.}
    \label{fig:GR_past_attractor}
\end{figure}

The analytic solutions $r(q)$ and $m(q)$ allow us to represent the cosmological solutions as a \emph{flow} in the theory space $r-m$, by means of a parametric plot. This is shown in Fig.\ref{fig:m_r_plotG1_analytical} corresponding to the same solutions that are represented by the trajectories in Fig.\ref{fig:phase_cosmo_G1}. Such a theory space representation of the solutions allows one to investigate whether GR behaves as a cosmological past or future attractor in the space of all $\Lambda$CDM-mimicking $f(Q)$ solutions. 

Lastly, for $\Gamma_1$, we have
\begin{equation}
        x_3 = -\frac{f_{QQ}\dot{Q}}{Hf_Q} = 12\dot{H}\frac{f_{QQ}}{f_Q} = 2(1+q)m =- (1+q) + \frac{3}{2}\Omega\,,
    \end{equation}
where at the last step, the trading equation \eqref{trading} has been used. This allows us to obtain
\begin{subequations}
\begin{align}
x_3(N) &= \frac{-6 C_2 e^{3N/2}}{\sqrt{2 - C_1 e^{3N}} \left( 2 - C_1 e^{3N} + 2 C_2 e^{3N/2} \sqrt{2 - C_1 e^{3N}} \right)}, 
\\
x_3(z) &= \frac{-6 C_2 (1 + z)^3}{\left[ 2(1 + z)^3 - C_1 \right] \left[ \sqrt{2(1 + z)^3 - C_1} + 2 C_2 \right]}.
\end{align}
\end{subequations}

The last of these expressions, namely the one of $x_3(z)$, is of particular interest, as it can be rewritten as a differential equation for the effective gravitational coupling by virtue of the definition of $x_3$:
\begin{equation}
\frac{d\ln(\kappa_{\rm eff})}{dz} = \frac{6 C_2 (1 + z)^2}{[2(1 + z)^3 - C_1] [\sqrt{2(1 + z)^3 - C_1} + 2 C_2]}\,.    
\end{equation}
The above equation can be solved analytically to obtain
\begin{equation}\label{ksec:f(Q)_cosmology_eff_G1}
   \kappa_{\rm eff}(z) = \frac{\sqrt{2(1 + z)^3 - C_1}}{\sqrt{2(1 + z)^3 - C_1} + 2 C_2},  
\end{equation}
where the integration constant is chosen such that $\kappa_{\rm eff}(z\to\infty)\to1$, i.e. on the basis of past asymptotic-ness to GR. Fig.\ref{fig:keff_c1} shows a typical evolution of the effective gravitational coupling.

\section{Connecting the model parameters with the present-day values of cosmological quantities}
\label{ss2c}

In this subsection, we aim to relate the model parameters of the $\Lambda$CDM-mimicking $f(Q)$ \eqref{flcdmG1} to actual cosmological quantities that are measurable, and provide possible bounds on them. For $\Gamma_1$, one has
    \beq 
    r = \frac{1}{x_1} = \frac{1}{2-\Omega} = \frac{1}{2-(\Omega_m/f_Q)} = \frac{1}{2-\kappa_{\rm eff}\Omega_m}\,.
    \eeq 
    Then one can calculate from \eqref{lambdac} that
    \beq 
    \Lambda_c(z) = \frac{(\beta \kappa_{\rm eff}\Omega_m)^2}{32 \alpha(\kappa_{\rm eff}\Omega_m - 1)}\,.
    \eeq 
    The above expression, along with the knowledge of the present-day values of the matter density abundance $\Omega_m$ and the effective gravitational coupling\footnote{Strictly speaking, the quantity $\kappa_{\rm eff}$ can be taken as an \emph{effective} gravitational coupling only in the subhorizon limit.} $\kappa_{\rm eff}$ allows one to obtain the following bounds on the model parameters for a real solution to exist (see the Table \ref{tab:lambda_branches})
    \beq\label{alphabetalambda_bound}
    \frac{\alpha\,\Lambda}{\beta^2} > \frac{\kappa_{\rm eff 0}^2\Omega_{m0}^2}{32(\kappa_{\rm eff 0}\Omega_{m0} - 1)}\,.
    \eeq 
By jointly using the definition of the effective gravitational coupling (\ref{kappaeff}), the $f(Q)$ theory (\ref{flcdmG1}), and the relationship between non-metricity and Hubble rate (\ref{Q-1}),  the previous inequality becomes the following implicit bound
\beq
\label{LL1A}
\frac{\alpha \Lambda}{\beta^2}>\frac{\rho_{m0}^2}{2(12\alpha H_0^2-\sqrt{6}\beta H_0-4\rho_{m0})(\sqrt{6}\beta-12\alpha H_0)H_0}\,,
\eeq
between the mathematical model parameters and the present-day values of relevant cosmological observables. Next, we note that by using Eq.(\ref{CC1}) together with (\ref{Omega_in_m}) and (\ref{trading}), we have\footnote{Within the $\Lambda$CDM-mimiking range (\ref{rangeq}) and positive $\alpha$, we correctly obtain negative $\Lambda$. }
\beq
\label{LL1}
\frac{\alpha\Lambda}{\beta^2} = \frac{3(1-2q_0)\Omega_0^2}{8[2(1+q_0) -3 \Omega_0]^2}\,, \qquad\qquad \Omega_0=\frac{\Omega_{m0}}{f_{Q0}}=\kappa_{\rm{eff}0}\Omega_{m0}\,.
\eeq
The above expression shows that the particular combination of model parameters that dictates the dynamics can essentially be translated into a combination of the present-day values of various cosmological quantities, and vice versa. Moreover, it also shows that at the background level, the vanishing of the cosmological constant term implies the vanishing of $\Omega_{m0}$. \footnote{This constitutes a refinement of the inequality \cite[Eq.(48)]{Chakraborty:2025qlv}.}  Eq.(\ref{LL1}) allows us to eliminate the model parameters on the left-hand side of (\ref{LL1A}) as\footnote{We note that we have removed an overall factor $(12\alpha H_0-\sqrt{6}\beta)$ which is positive because of the requirement $f_Q>0.$}
\beq 
\frac{3(1-2q_0)}{[(1+q_0)(12\alpha H_0-\sqrt{6}\beta)H_0-6 \rho_{m0}]^2}-\frac{1}{4\rho_{m0} - (12\alpha H_0-\sqrt{6}\beta)H_0} >0\,.
\eeq 
Finally, the present-day density of matter can be replaced by the adimensional matter abundance parameter leading to
\beq
\frac{(\beta-2\sqrt{6}\alpha H_0)H_0[2(1+q_0)-3\Omega_0]^2}{\Omega_0-1}-2\sqrt{6}(q_0-1)>0\,.
\eeq

An independent route for finding the redshift evolution of the effective matter density abundance parameter consists of enforcing the dust conservation law and the $\Lambda$CDM-mimicking Hubble rate into (\ref{dynvar_gamma1}). Via this route, we obtain:
\beq
\Omega (z)=\frac{2\sqrt{6}\,\rho _{m0}(1+z)^{3}}{3H_{0}h(z)\left(2\sqrt{6}\,\alpha\,H_{0}h(z)-\beta \right)}\,, \qquad\qquad h^2(z)=\frac{2}{3}(1+q_0)(1+z)^3 + \frac{1}{3}(1-2q_0)\,.
\eeq
It is interesting to note that, although the nonvanishing $\beta\sqrt{-Q}$ term does not affect the cosmological \emph{kinematics}, i.e. the evolution of the scale factor $a(t)$ (see Appendix \ref{appA}), it does however has a \emph{dynamical} effect, in the sense that it affects the evolution of the matter abundance.

At the present cosmic epoch, we have 
\beq
\Omega _{0}=\frac{2\sqrt{6}\rho _{m0}}{3H_{0}\left(2\sqrt{6}\,\alpha\,H_{0}-\beta \right)}\,,
\eeq
which sets an upper bound on the model parameter combination $\frac{\beta}{\alpha}$:
\be\label{beta_bound}
\frac{\beta}{\alpha}<2\sqrt{6}\,H_0\,.
\ee

Taking into consideration all the possible bounds that we have obtained so far, namely Eqs.\eqref{lambda_bound},\eqref{alpha_bound},\eqref{alphabetalambda_bound} and \eqref{beta_bound}, we can list the following conditions on the model parameters $\alpha,\beta,\Lambda$ for a physically realistic late-time $\Lambda$CDM-like cosmological solution
\be
\Big\{(\alpha,\beta,\Lambda):\,\alpha>0,\,\,\Lambda>0,\,\,\beta<2\sqrt{6}H_0\alpha,\,\,\frac{\Lambda/\alpha}{(\beta/\alpha)^2} > \frac{\kappa_{\rm eff 0}^2\Omega_{m0}^2}{32(\kappa_{\rm eff 0}\Omega_{m0} - 1)}\Big\}\,.
\ee 

Without the boundary term $\beta=0$, a constraint on the effective coupling directly translates into a constraint on $\alpha$ (the same result characterizes the GR submanifold $m=0$). Then, we can note that we have a neat separation between a  GR + correction term: $(\kappa_{\rm{eff}0})^{-1}=\frac{12\alpha H_0 -\sqrt{6}\beta}{12 H_0}= (\kappa_{\rm{GR}0})^{-1}-\frac{\sqrt{6}\beta}{12 H_0}$: this confirms that the boundary term is indeed relevant when we study perturbation because it supports the deviation from GR. Thanks to the third condition listed above, the effective coupling is positive as it should be. More quantitatively: we pin down $H_0$ with astrophysical data at the background level, then, to tame the $\sigma_8$ tension, do we need a smaller effective coupling? This should give the range or algebraic sign of $\beta$. This is a hot topic of discussion because the boundary term is exactly what suppresses the formation of astrophysical structures \cite{Li:2025msm}.

The matching with (\ref{OOR1}) at every redshift delivers the following relationships between the free model parameters:
\beq
\label{astroG1}
\Omega_{m0}\equiv\frac{\rho_{m0}}{3H_0^2}=\frac{2}{3}(1+q_0)\alpha\,, \qquad \Omega_{\Lambda0}\equiv\frac{\Lambda}{3H_0^2}=\frac{1}{3}(1-2q_0)\alpha\,.
\eeq

\section{Robustness of the cosmological dynamics against slight deviation from the kinematic condition $j=1$}\label{sec:robustness}

So far, we have considered only the exact $\Lambda$CDM-like evolution $j=1$. This, of course, is an idealization, as a realistic cosmic evolution is most possibly one that is not exactly $j=1$ but slightly away from it. In this section, let us consider a family of "almost $\Lambda$CDM models", motivated by the recent works \cite{Chakraborty:2025rvc,Worsley:2026ijo}:
\beq \label{almost_LCDM}
j=1+\epsilon(q-q_*)\,.
\eeq
Here $q_*$ is a value of $q$ that acts as a ``pivot'', which determines at what epoch the cosmic evolution matches the $\Lambda$CDM-like cosmic evolution asymptotically. $q_*=1/2,-1$ corresponds to the cosmic evolutions becoming asymptotically $\Lambda$CDM-like in the asymptotic past and future respectively.

The dynamical system \eqref{evcosmoG1} now reads
\begin{subequations}
\label{sysrobust}
	\begin{eqnarray}
		\Omega' &=& \frac{3}{2}\Omega\left[\Omega-\frac{2}{3}(2-q)\right]\,,\\
		q' &=& 2 q^2+(1-\epsilon)q-1+\epsilon_1q_* = 2\left(q - \frac{\epsilon-1+2\sigma}{4}\right)\left(q - \frac{\epsilon-1-2\sigma}{4}\right)\,,
	\end{eqnarray}
\end{subequations}
with
\begin{equation}
\label{defsigma}
    \sigma=\frac{\sqrt{9+\epsilon^2-2(4q_*+1)\epsilon}}{2}\,.
\end{equation}

\begin{figure}[H]
\centering
	\subfigure[]{%
		\includegraphics[width=8cm,height=6cm]{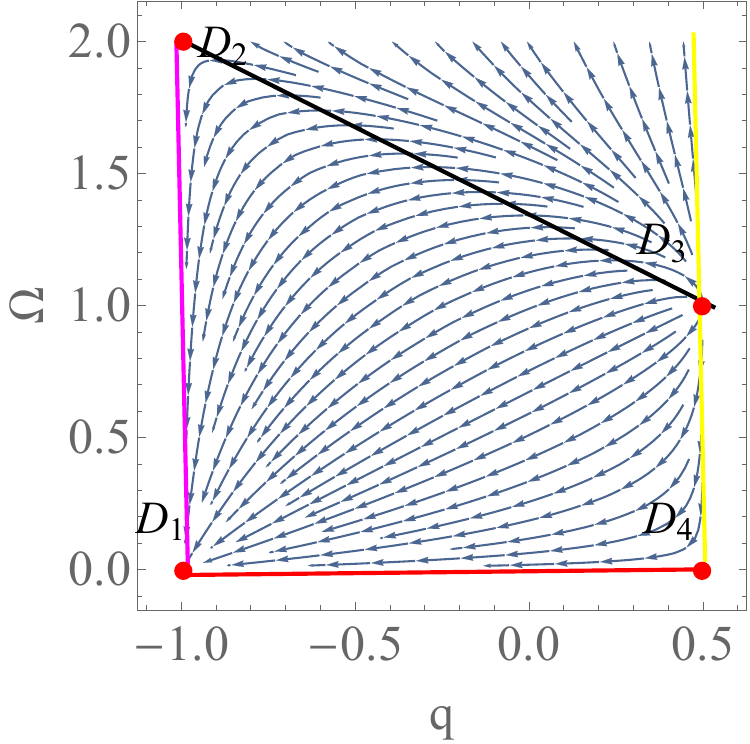}\label{fig:phport_eps1}}
	\quad
	\subfigure[]{%
		\includegraphics[width=8cm,height=6cm]{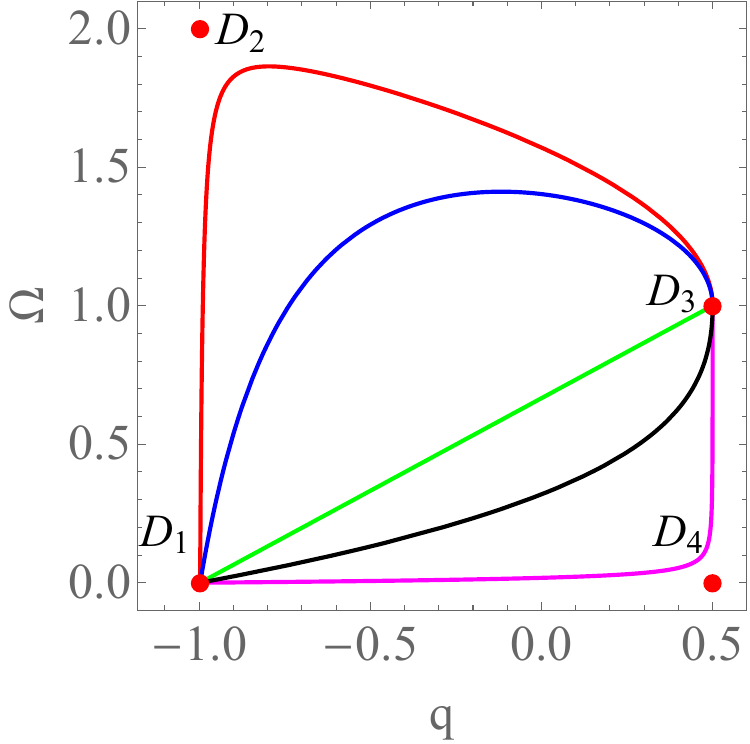}\label{fig:phsoln_eps1}}
	\caption{Phse portrait and the solution curves corresponding to $j=1+\epsilon(q-1/2)$ with a typical small value of th deviation parameter $\epsilon=0.006$. Compare with Fig.\ref{fig:phase_cosmo_G1} for the $j=1$ case.} 
\label{fig:robust_eps1}
\end{figure}

\begin{figure}[H]
\centering
	\subfigure[]{%
		\includegraphics[width=8cm,height=6cm]{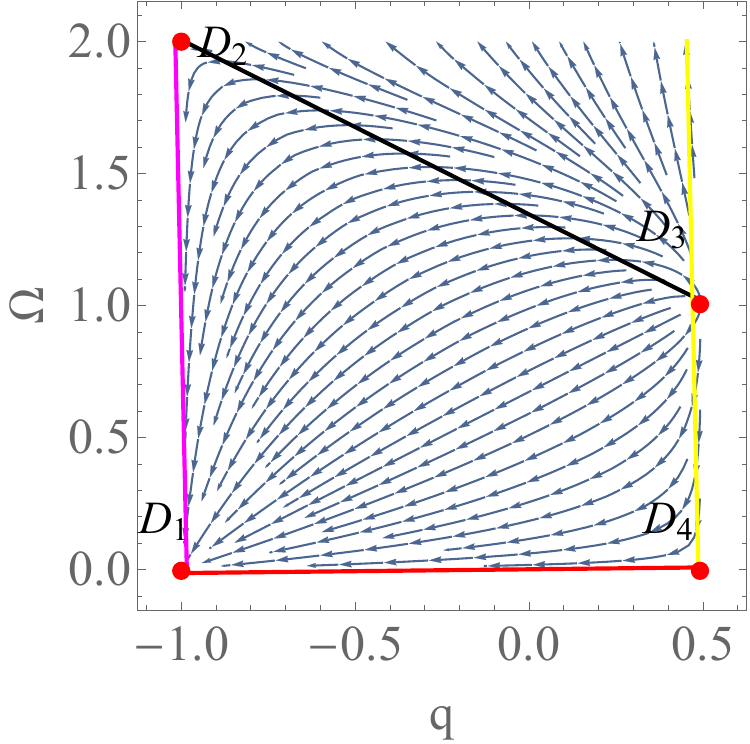}\label{fig:phport_eps2}}
	\quad
	\subfigure[]{%
		\includegraphics[width=8cm,height=6cm]{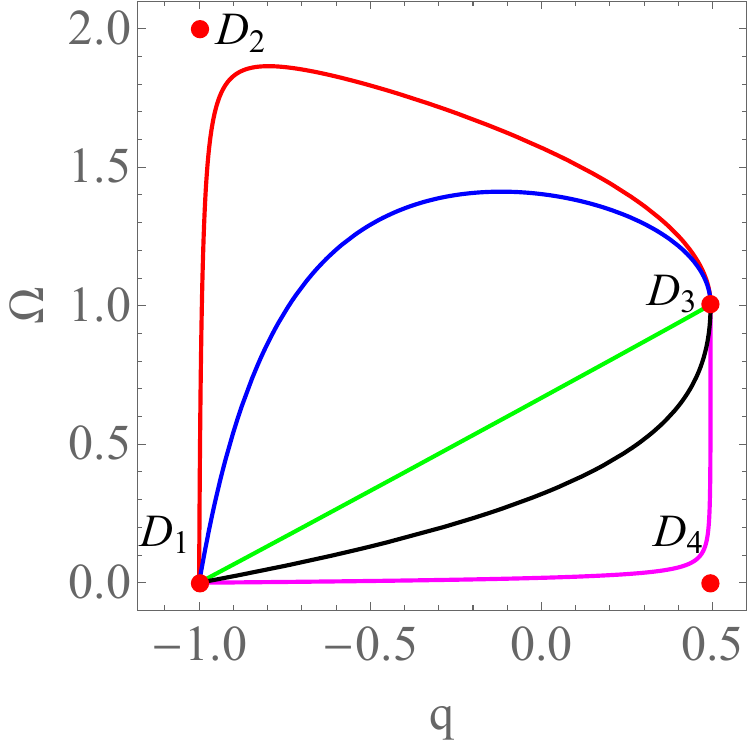}\label{fig:phsoln_eps2}}
	\caption{Phase portrait and the solution curves corresponding to $j=1+\epsilon(q+1)$ with a typical small value of the deviation parameter $\epsilon=-0.012$. Compare with Fig.\ref{fig:phase_cosmo_G1} for the $j=1$ case.} 
    \label{fig:robust_eps2}
\end{figure}

The system \eqref{sysrobust} presents the invariant submanifold $\Omega=0$, as well as two invariant submanifolds in the $q$ direction given by $q=\frac{\epsilon-1 \pm 2\sigma}{4}$. Within the interval $\frac{\epsilon-1 - 2\sigma}{4} < q < \frac{\epsilon-1 + 2\sigma}{4}$, $q$ is a monotonically decreasing function. These features are similar to the $j=1$ case. In fact, for $\epsilon=0$, the two invariant submanifolds in the $q$-direction reduces to $q=\frac{1}{2},-1$. Therefore, one expects the phase portrait to be similar to that in Fig.\ref{fig:phase_cosmo_G1}, with the left and right $q$-boundaries slightly shifted. This is confirmed in Figs.\ref{fig:robust_eps1} and \ref{fig:robust_eps2}. 

The fixed points of the system \eqref{sysrobust} are listed in Table \ref{tablerobust}. One can check that the fixed points $D_i$'s have a one-to-one correspondence with the fixed points $B_i$'s of Table \ref{tab:G1_cosmo_pts}, to which they reduce in the limit $\epsilon\to0$. 
\begin{table}[H]
		\centering
	\renewcommand{\arraystretch}{1.6}
		\begin{tabular}{|c|c|c|c|c|c|c|}
			\hline 
			Fixed point  & $(\Omega, q)$ & j & Stability  &  Cosmological solution & Eigenvalues \\ 
			\hline 
			$D_1$ & $\left(0,\, \frac{\epsilon-1-2\sigma}{4} \right)$ & $j=1+\frac{\epsilon[\epsilon-1-2(\sigma+2q_*)]}{4}$ &  Attractor & $a(t)=a_0 t^{\frac{4}{3+\epsilon-2\sigma}}$  & $\left(-\sigma,\, \frac{\epsilon-9-\sigma}{4} \right)$ \\
			\hline			
			$D_2$ & $\left(\frac{9-\epsilon+2\sigma}{6},\, \frac{\epsilon-1-2\sigma}{4} \right)$ & $j=1+\frac{\epsilon[\epsilon-1-2(\sigma+2q_*)]}{4}$ & Saddle & $a(t)=a_0 t^{\frac{4}{3+\epsilon-2\sigma}}$ & $\left(-\sigma,\, \frac{9-\epsilon+\sigma}{4} \right)$\\
			\hline 
			$D_3$ & $\left(\frac{9-\epsilon-2\sigma}{6},\, \frac{\epsilon-1+2\sigma}{4} \right)$ & $j=1+\frac{\epsilon[\epsilon-1+2(\sigma-2q_*)]}{4}$ & Repeller & $a(t)=a_0 t^{\frac{4}{3+\epsilon+2\sigma}}$ &$\left(\sigma,\, \frac{9-\epsilon-\sigma}{4} \right)$\\ 
            \hline	
			$D_4 $ & $\left(0,\, \frac{\epsilon-1+2\sigma}{4} \right)$ & $j=1+\frac{\epsilon[\epsilon-1+2(\sigma-2q_*)]}{4}$ & Saddle & $a(t)=a_0 t^{\frac{4}{3+\epsilon+2\sigma}}$ & $\left(\sigma,\, \frac{\epsilon-9+\sigma}{4} \right)$\\ 		
			\hline 		
		\end{tabular}
		\caption{In this table, we list the fixed points for the system \eqref{sysrobust} together with their stability and the cosmology they support. We denote them with $D$ rather than $C$ to prevent notational confusion with the arbitrary integration constants arising in some unrelated equations. For enlightening the unifying structure of the eigenvalues, we have introduced the notation $\sigma=\sqrt{9 + \epsilon (\epsilon-2- 8 q_*)}/2$.}
		\label{tablerobust}
	\end{table} 

One can notice that as the attracting invariant submanifold slightly shifts from the value $q=-1$, the two fixed points $D_1$ and $D_2$ are not de-Sitter solutions anymore. However, for small values of the deviation parameter $\epsilon$, these fixed points still indicate an accelerating solution. Overall,
the above analysis indicates that the phase portrait of the $\Lambda$CDM-mimicking $f(Q)$ cosmology is structurally stable, i.e., retains its essential qualitative feature, against small deviations from the exact $\Lambda$CDM-kinetics.

It turns out that, just like for the exact $\Lambda$CDM background $j=1$, even for the almost $\Lambda$CDM background $j=1+\epsilon(q-q_*)$, the system is integrable; although the expressions are undoubtedly more complicated. Following what was done in Sec.\ref{subsec:analytic}, one can integrate the evolution of the deceleration function in a closed form as
\ba 
\label{qNrobust}
q(N) &=& -\frac{\sigma}{2}\tanh\left[\sigma (N -N_0)- {\rm arctanh}\left( \frac{1+4q_0-\epsilon}{2 \sigma}\right)\right]+\frac{\epsilon-1}{4}\,, 
\\
q(z) &=& -\frac{\sigma}{2} \tanh \left[\sigma \ln \left(\frac{1+z_0}{1+z} \right) - {\rm arctanh}\left( \frac{1+4q_0-\epsilon}{2 \sigma}\right)\right]+\frac{\epsilon-1}{4}\,.
\ea

Next, we obtain
\begin{eqnarray}
\Omega(N) &=& \frac{(13-\epsilon-3\sigma)e^{(\epsilon-13)N/6}}{(13-\epsilon-3\sigma){\mathcal D}_2 \sqrt{\cosh(\sigma N+{\mathcal D}_1 )}+9e^{(\epsilon-13)N/6}(1+e^{2(\sigma N+{\mathcal D}_1 )}) \,_2F_1\left[1,\, \frac{\epsilon+9\sigma-13}{12 \sigma},\,\frac{\epsilon+15\sigma-13}{12 \sigma},\,-e^{2(\sigma N+{\mathcal D}_1 )}  \right]} \,, \nonumber\\
&&
\end{eqnarray}
with ${\mathcal D}_1$ implicitly defined by the transcendental condition $\Omega(N_0)=\Omega_0$ and
\begin{equation}
    {\mathcal D}_2 = -\left[\sigma N_0+ {\rm arctanh}\left( \frac{1+4q_0-\epsilon}{2 \sigma}\right)\right]\,,
\end{equation}
and $_2F_1$ being one of the Hypergeometric functions. Hence,
\begin{equation}
\Omega(q) = \frac{(13-\epsilon-3\sigma)(1+4q-\epsilon+2\sigma)
[4 \sigma^2-(\epsilon-1-4q)^2]^{1/4}e^{\frac{(13-\epsilon){\mathcal D}_1}{6\sigma}}\left(\frac{1+4q-2\sigma-\epsilon}{2\sigma+1+4q-\epsilon} \right)^{\frac{13-\epsilon}{12 \sigma}}}{{\mathcal K}_1(q) +{\mathcal K}_2(q)}\,,    
\end{equation}
with
\begin{eqnarray}
{\mathcal K}_1(q) &=& {\mathcal D}_2 \sqrt{2\sigma}
(13-\epsilon-3 \sigma)(1+4q-\epsilon1+2 \sigma)\,, \\
{\mathcal K}_2(q)&=& 36 \sigma [4 \sigma^2-(\epsilon-1-4q)^2]^{1/4}e^{\frac{(13-\epsilon){\mathcal D}_1}{6\sigma}}\left(\frac{1+4q-2\sigma-\epsilon}{2\sigma+1+4q-\epsilon} \right)^{\frac{13-\epsilon}{12 \sigma}} \\ \nonumber
&& \qquad\cdot\,\,_2F_1 \left[1,\, \frac{\epsilon+9\sigma-13}{12 \sigma},\,\frac{\epsilon+15\sigma-13}{12 \sigma},\,\frac{2\sigma-1-4q+\epsilon}{2\sigma+1+4q-\epsilon} \right]\,.
\end{eqnarray}

It must be appreciated that the cosmography-driven closure of the dynamical system allows us to establish, in a rather straightforward manner, the robustness of the underlying $f(Q)$ cosmological dynamics against slight deviation from the exact $\Lambda$CDM-like cosmic evolution, completely bypassing the actual reconstruction of the functional form of $f(Q)$. Although using the trading equation \eqref{trading} to calculate $m=\frac{Qf_{QQ}}{f_Q}$ at a particular fixed point, we can reconstruct the functional form of $f(Q)$ in the vicinity of that particular fixed point, the reconstruction of $f(Q)$ for an entire evolution of the form $j=1+\epsilon(q-q_*)$ would hit algebraic roadblock due to invertibity issues; the reader can revisit the reconstruction method in \cite{Chakraborty:2025qlv}. Anyway, the cosmographic closure method for the dynamical system renders that cumbersome excercise unnecessary.

\section{Stability of the $\Lambda$CDM-like cosmological solution}\label{sec:stability}

In general, a cosmological solution $H(t)$ is characterized by a phase trajectory in the phase space, whereas fixed points characterize the asymptotic states of the solutions that are interpreted as cosmological epochs. For example, in Fig.\ref{fig:phase_cosmo_G1}, all the trajectories flowing from $B_3$ to $\mathcal{B}_1$ are possible $\Lambda$CDM-like ($j(z)=1$) $f(Q)$ cosmological solutions, whereas the fixed points $B_3$ and $B_1$ signify matter-dominated past epoch and dark energy-dominated future epoch, respectively. Following the idea introduced in \cite{Chakraborty:2025qlv} (in particular, see \cite[Sec. 6.4]{Chakraborty:2025qlv}), we analyze the stability of $\Lambda$CDM-like cosmological solutions in $f(Q)$ gravity against small homogeneous and isotropic perturbations, from the point of view of the cosmographic phase space approach.

Eq.\eqref{rayG1} can be written as
\be
2\dot{H} f_Q - 24\dot{H}H^2 f_{QQ} + 6H^2 f_Q + \frac{1}{2}f = 0\,.
\ee
Consider any given solution $H(t)$ of this equation, and take a homogeneous and isotropic perturbation around it: $H(t)\rightarrow H(t)+\delta H(t)$. Substituting it back into the above equation and keeping terms of only linear order, we get (see \cite[Eq.(6.25)]{Chakraborty:2025qlv}) 
\be\label{stability}
\frac{d\delta H}{dN} + \lambda(Q) \delta H = 0\,, \qquad \lambda(Q) = 3 - 2(1+q)\frac{(3 + 2 m_2)m}{1+2m}\,,
\ee
where $m_2$ is the next order parameter after $m$ introduced in the hierarchy \eqref{hierarchy}. Combining the recursion relation \eqref{recursion} and the trading equation \eqref{trading}, one can show that it is possible to express $m_2$ as
\begin{equation}\label{m2_f(Q)}
    m_2 = -\frac{3}{2}-\frac{(j-1)}{2(1+q)^2}\left[\frac{\Omega}{\Omega-\frac{2}{3}(1+q)}\right]\,.
\end{equation}
When the expression of $m$ from Eq.\eqref{trading} and the expression of $m_2$ from Eq.\eqref{m2_f(Q)} is substituted, the quantity $\lambda(Q)$ takes a remarkably nice and simple form 
\begin{equation}\label{lambda}
    \lambda(Q) = \frac{j+3q+2}{q+1}\,,
\end{equation}
which lets us investigate the stability with respect to small homogeneous and isotropic perturbations of any given cosmological solution $j=j(q)$ in $f(Q)$ gravity. 

It can be appreciated that the behaviour of homogeneous isotropic small perturbations is fully determined by the cosmographic parameters $q$ and $j$. This is a consequence of the field equations \eqref{fieldG1} being second order in the metric, likewise in GR.

In particular, for a $\Lambda$CDM-like cosmological solution, $j=1$, one can see that $\lambda(Q)=3$ identically, leading to
\beq 
\frac{d\delta H}{dN} + 3\delta H = 0 \quad \Rightarrow \delta H \sim e^{-3N} \sim a^{-3}\,.
\eeq
Small homogeneous and isotropic perturbations around a $\Lambda$CDM-like cosmological solution decay as $\sim a^{-3}$, implying that a $\Lambda$CDM-like solution in the underlying $f(Q)$ is a stable solution.

We can, in fact, also investigate whether an almost $\Lambda$CDM like cosmological solution, like in Eq.\eqref{almost_LCDM}, can arise as a stable solution in the $f(Q)$ framework. Substituting $j=1+\epsilon(q-q_*)$ in the expression \eqref{lambda}, we get
\beq 
\lambda(Q) = 3 + \frac{\epsilon(q-q_*)}{q+1}\,.
\eeq 
For small values of $\epsilon$, i.e., small deviation from the exact kinematics of $\Lambda$CDM, $\lambda(Q)>0$, implying that the respective cosmological solution is still stable with respect to homogeneous and isotropic perturbations. 

The stability of a cosmological solution against small homogeneous and isotropic perturbations is related to whether or not achieving such a solution requires fine-tuned initial conditions; see \cite[Sec. 6]{Chakraborty:2025lkz} for a nice discussion about it. The stability of the $\Lambda$CDM-like cosmological solution in $f(Q)$ implies that it does not require a fine-tuned initial condition for such a solution to be realized in the $f(Q)$ framework.

\section{Matter perturbation evolution}\label{sec:ptbn}

As of now, we have considered a $\Lambda$CDM-mimicking $f(Q)$ model, namely, the $f(Q)$ theory that reproduces exactly the $\Lambda$CDM-like cosmic evolution, as given by the cosmographic condition $j(z)=1$. Evidently, such a model is indistinguishable from the standard $\Lambda$CDM model of cosmology at the background level. However, differences will show up at the perturbation level, because of the existence of the additional $\sqrt{-Q}$ term making the theory nonequivalent to GR. The perturbative signature of the $\sqrt{-Q}$ term in the action has recently been investigated in \cite{Li:2025msm}. Along the same line, we consider the linear evolution of matter density in this section, along with the growth index parameter. As we will show below, the cosmographic closure approach allows us to obtain analytical solutions even in the perturbation sector. Thus, we provide a rare example in modified gravity in which both the background dynamics and the linear matter perturbations both admit closed-form expressions. We can use this solution to compare with observational probes without relying entirely on numerical integrations.

We will achieve so by taking advantage of the  progress on the application of the Heun  function to perturbation problems of gravitational systems \cite{Minucci:2024qrn}, which has led to the implementation of such a special function into software for algebraic and symbolic manipulations only recently  \cite{Chen:2023ese}. In fact,   the differential equation obeyed by the redshift-rescaled logarithmic derivative of the density contrast will be shown to be of the  Riccati type, which can be cast into the Heun form via a suitable transformation of variable involving the ratio of the Heun function to its first derivative. In black hole physics, the Teukolsky equation serves as the master equation governing radial perturbation evolution. Its solutions, expressed via Heun functions, allow the extraction of frequency and amplitude spectra for gravitational waves and quasinormal modes. While black hole perturbation theory sets boundary conditions for wavefunction decay at the horizon and spatial infinity, our spatially homogeneous cosmological model instead imposes boundary conditions by matching with General Relativity in the far past during the matter-dominated era.

In the  connection branch $\Gamma_1$, the matter density contrast $\delta_m=\frac{\delta\rho_m}{\rho_m}$ in the linear approximation at \emph{sub-horizon scales} follows an equation which can be expressed in the following three equivalent forms 
\begin{subequations}
\begin{align}
& \frac{d^2 \delta_m}{da^2}+\left(\frac{dH}{H da} +\frac{3}{a}\right)\frac{d \delta_m}{da}-\frac{3 \Omega_{m0}}{2 H^2 a^5}\frac{\kappa_{\rm eff}}{\kappa}\delta_m = 0\,,
\\
& \frac{d^2 \delta_m}{dz^2}  +\left( \frac{dH}{Hdz}-\frac{1}{1+z}\right) \frac{d \delta_m}{dz}-\frac{3 \Omega_{m0}(1+z)}{2 H^2 }\frac{\kappa_{\rm eff}}{\kappa}\delta_m = 0\,,
\\
& \frac{d^2 \delta_m}{dN^2} + (1 - q) \frac{d \delta_m}{dN} - \frac{3 \Omega_{m0}}{2 H^2 e^{3N}} \frac{\kappa_{\rm eff}}{\kappa} \delta_m = 0\,.
\end{align}
\end{subequations}
The first of the above three forms appears in \cite[Eq.(7)]{Anagnostopoulos:2021ydo}.

Defining the quantity $U_m \equiv \frac{d \delta_m}{\delta_m dN}$, the evolution equation can be expressed as a first-order equation for $U_m$, which can be expressed in three equivalent forms
\begin{subequations}
\begin{align}
& \frac{dU_m}{dN} + U_m^2 + (1-q)U_m - (1+q)(1+2m) = 0  \,,
\\
& \frac{dU_m}{dz} - \frac{U_m^2}{1+z} + \left( \frac{dH}{Hdz} - \frac{2}{1+z} \right)U_m + \frac{3 \Omega_{m0} (1+z)^2}{2 H^2} \frac{\kappa_{\text{eff}}}{\kappa}=0 \,,
\\
& \frac{dU_m}{dq} = \frac{(1+q)(1+2m) - (1-q)U_m - U_m^2}{2q^2+q - 1}\,.
\end{align}
\end{subequations}
The first equation is natural in terms of the e-folding number $N$, while the second is directly suitable for comparison with observations through the redshift $z$. While all three equations belong to the Riccati class, the third is particularly interesting. This is because, upon expressing $m$ as a function of $q$, it can be reduced to a second-order non-linear differential equation. In the present case, the resulting equation is identified as the general Heun equation. Consequently, the growth rate admits the following exact analytical solution in terms of the general Heun functions and their derivatives:

\begin{eqnarray}\label{Um_soln}
    U_m(q) = \frac{2(1+q)^2 \Big[ A(q)H_1'(\zeta(q)) + B(q)H_1(\zeta(q)) + C(q)H_2(\zeta(q)) + D(q)H_2'(\zeta(q)) \Big]}{X^3 y^{2/3} \left(\sqrt{y}+\sqrt{3}\right) \Big[ X^{5/3}H_2(\zeta(q)) + \mathcal{C}_1 y^{5/6}H_1(\zeta(q)) \Big]}\,,
\end{eqnarray}
where $\mathcal{C}_1$ is a constant of integration and
\begin{eqnarray}
y&=& 1-2q,~~X=\sqrt{3}-\sqrt{y}\,,
\\ 
\tilde{\mathcal C}&=&\frac{2(1+q_0)-3\Omega_0}{\sqrt{1-2q_0}\Omega_0}\,,
\\
q_1 &=& -\frac{ \left((\tilde{\mathcal{C}}^3+19\tilde{\mathcal{C}})\sqrt{3}-14\tilde{\mathcal{C}}^2-24\right)}{9(\tilde{\mathcal{C}}-\sqrt{3})^2}\,,
\\
q_2 &=& -\frac{\sqrt{3}}{9}\left(\tilde{\mathcal{C}}-\sqrt{3}\right)\,,
\\
A(q) & =& -(\tilde{\mathcal{C}}+\sqrt{3})\mathcal{C}_1 y^2\,,
\\
B(q) &=& \frac{2}{3}\mathcal{C}_1 y^{3/2} \left(3 - \sqrt{3y}\right)\,,
\\
C(q) &=& -\sqrt{3} y^{2/3} X^{8/3}\,,
\\
D(q) &=&-\frac{1}{2} \left(\tilde{\mathcal{C}}+\sqrt{3}\right) y^{7/6} X^{8/3}\left(\sqrt{y}+\sqrt{3}\right)\,,
\\
\zeta(q) &=& -\frac{(\tilde{\mathcal{C}}+\sqrt{3})\sqrt{y}}{\sqrt{3}X}\,,
\end{eqnarray}
and HeunG, HeunGPrime are the general Heun function and its first derivative with respect to the argument $\zeta(q)$:
\begin{eqnarray}
     H_1(\zeta(q))& =& \text{HeunG}\left(\frac{1}{2} + \frac{\sqrt{3}}{6}\tilde{\mathcal{C}}, \, q_1, \, \frac{2}{3}, \, \frac{4}{3}, \, \frac{8}{3}, \, 0, \, \zeta(q)\right)\,, \\
    H_1'(\zeta(q)) &=& \text{HeunGPrime}\left(\frac{1}{2} + \frac{\sqrt{3}}{6}\tilde{\mathcal{C}}, \, q_1, \, \frac{2}{3}, \, \frac{4}{3}, \, \frac{8}{3}, \, 0, \, \zeta(q)\right)\,, \\
    H_2(\zeta(q)) &=& \text{HeunG}\left(\frac{1}{2} + \frac{\sqrt{3}}{6}\tilde{\mathcal{C}}, \, q_2, \, -1, \, -\frac{1}{3}, \, -\frac{2}{3}, \, 0, \, \zeta(q)\right)\,, \\
    H_2'(\zeta(q)) &=& \text{HeunGPrime}\left(\frac{1}{2} + \frac{\sqrt{3}}{6}\tilde{\mathcal{C}}, \, q_2, \, -1, \, -\frac{1}{3}, \, -\frac{2}{3}, \, 0, \, \zeta(q)\right) \,.
\end{eqnarray}
In the above, the constant $\tilde{\mathcal{C}}$ is the same constant that appears in Eq.\eqref{Omegavsq}. On the other hand, $\mathcal C_1$ is the integration constant associated with the perturbation sector. As anticipated above, the appearance of the general Heun functions reflects the mathematical structure of the underlying perturbation equation rather than a special ansatz. 

It is useful to note that from Eqs.~ \eqref{qlcdm} and \eqref{HLCDM}, we indeed have
\begin{equation}
y=1-2q
=\frac{3(1-c_1)}{1-c_1+c_1(1+z)^3}
=\frac{3(1-c_1)H_0^2}{H^2(z)}.
\end{equation}
 Thus, the auxiliary quantity $y$ appearing in the Heun solution is directly determined by the background expansion history.

Since the solutions $q(N)$ or $q(z)$ are already known for a $\Lambda$CDM-like evolution (Eq.\eqref{qlcdm}), substitution of those expressions into the solution \eqref{Um_soln} allows for $U_m$ to be expressed as a function of either $N$ or $z$. A quantity of particular observational interest is the growth index parameter $\gamma$,  defined via the relation
\be\label{growth_function}
U_m = \Omega_m^{\gamma} = \left(\frac{\Omega}{\kappa_{\rm eff}}\right)^{\gamma}\,.
\ee 

For the standard GR $\Lambda$CDM model, $\gamma\approx6/11\approx0.55$. However, for $\Lambda$CDM-mimicking modified gravity models, not only is $\gamma$ different, but demonstrates a nontrivial time evolution: $\gamma=\gamma(z)$ (see, e.g., \cite{Samanta:2026pvb}).  Consequently, the evolution of $\gamma$ provides a useful discriminator between GR and modified gravity models. Before we can make a quantitative comparison between the models, we have to pin down the integration constant $\mathcal{C}_1$ in the analytical solution $U_m(q)$ of Eq.\eqref{Um_soln} through an appropriate physical initial condition. 

Physically, GR is expected to provide an excellent description of the Universe during the matter-dominated era. Accordingly, a viable $\Lambda$CDM-mimicking $f(Q)$ model should asymptotically recover the standard GR $\Lambda$CDM behaviour at sufficiently high redshift. We therefore impose the initial condition $\gamma(q\approx0.49)=6/11$, 
which corresponds to the well-known growth index of the GR $\Lambda$CDM model near the matter-dominated epoch.

The analytical solution for the effective matter abundance is already known from Eq.~\eqref{Omegavsq},
\be 
\Omega(q)=\frac{2(1+q)}{3 + \tilde{\mathcal C}\sqrt{1-2q}}\,.
\ee
The analytic solution $\kappa_{\rm eff}(q)$ can be obtained by combining the expression of $q(z)$ from \eqref{qlcdm} with the solution $\kappa_{\rm eff}(z)$ from \eqref{ksec:f(Q)_cosmology_eff_G1}:
\be 
\kappa_{\rm eff}(q)=\frac{1}{1+\frac{\tilde{\mathcal{C}}}{3}\sqrt{1-2q}}\,.
\ee 
Using the defining relation \eqref{growth_function}, and the analytical solutions of $\Omega(q),\,\kappa_{\rm eff}(q)$, we can write 
\begin{equation}
U_m=\left(\frac{\Omega}{\kappa_{\rm eff}}\right)^\gamma =
\left[\frac{\displaystyle \frac{2(1+q)}{3+\tilde{\mathcal C}\sqrt{1-2q}}}
{\displaystyle \frac{1}{1+\frac{\tilde{\mathcal C}}{3}\sqrt{1-2q}}}\right]^{\gamma}
=
\left[\frac{2}{3}(1+q)\right]^{\gamma}\,.
\end{equation}
Then, the aforementioned initial condition yields
\begin{equation}
U_m(q\simeq0.49)=
\left(\frac{2}{3}\times1.49\right)^{6/11}.
\end{equation}

The above equation serves to pin down the integration constant $\mathcal{C}_1$ in the solution \eqref{Um_soln}, which comes out as 
\begin{equation}
 \mathcal{C}_1=   \frac{\left( 333.94 \tilde{\mathcal{C}} + 578.40 \right) H_1' + 10500.57 H_1}{ 6.06 H_2- \left( 5.91 \tilde{\mathcal{C}} + 10.24 \right) H_2' }\,,
\end{equation}
where
\begin{align*}
H_1  &= \text{HeunG}\left( \alpha, \chi_1, -1, -\frac{1}{3}, -\frac{2}{3}, 0, \beta \right) \,,
\\
H_1' &= \text{HeunGPrime}\left( \alpha, \chi_1, -1, -\frac{1}{3}, -\frac{2}{3}, 0, \beta \right)\,, 
\\
H_2  &= \text{HeunG}\left( \alpha, \chi_1 \chi_2, \frac{2}{3}, \frac{4}{3}, \frac{8}{3}, 0, \beta \right)\,,
\\    
H_2' &= \text{HeunGPrime}\left( \alpha, \chi_1 \chi_2, \frac{2}{3}, \frac{4}{3}, \frac{8}{3}, 0, \beta \right)\,,
\\
\alpha &= 0.50 + 0.29 \tilde{\mathcal{C}}, \quad \beta = -0.05 \tilde{\mathcal{C}} - 0.09 \,,
\\
\chi_1 &= \frac{(\tilde{\mathcal{C}}^2 - 3)^2}{\left(1.73\tilde{\mathcal{C}}+ 3\right)\left(1.73 \tilde{\mathcal{C}} - 3\right)^2}\,, \qquad \qquad
\chi_2 =1.73\tilde{\mathcal{C}}^3 + 2\tilde{\mathcal{C}}^2 -22.52\tilde{\mathcal{C}} + 24\,.
\end{align*}

This value of $\mathcal{C}_1$ guarantees that deep inside the matter domination, even linear perturbations behave the same way as in GR, as they should. Perturbative deviations become apparent only at later times, where they provide a potential observational signature of the model. 

After fully determining the analytical solution, we can now directly evaluate the key quantity characterizing the growth of the large-scale structure. In particular, the growth index parameter can be written as
\begin{equation}
    \gamma(z) = \ln U_m(z)/\ln\left(\frac{\Omega(z)}{\kappa_{\rm eff}(z)}\right) = \ln U_m(z)/\ln\left(\frac{2}{3}(1+q)\right)\,.
\end{equation}
In the numerator of the above expression, $U_m(z)$ is found by substituting the expression $q(z)$ from Eq.\eqref{qlcdm} into the solution $U_m(q)$. As for the denominator, using the analytical expression for $q(z)$ from Eq.~\eqref{qlcdm}, together with Eq.~\eqref{HLCDM}, one can write
\begin{equation}
\frac{2(1+q)}{3}
=
\frac{c_1(1+z)^3}{c_1(1+z)^3 + (1-c_1)}\,, \qquad\qquad c_1=\frac{2}{3}(1+q_0).
\end{equation}
Hence, the growth index takes the compact form 
\begin{equation}
\label{finalgamma}
\gamma(z)
=\frac{\ln U_m(z)}
{\ln\!\left(\frac{c_1(1+z)^3}{c_1(1+z)^3 + (1-c_1)}\right)}\,, \qquad\qquad c_1=\frac{2}{3}(1+q_0).
\end{equation}

The above expression separates the background and perturbation contributions nicely. While the denominator is completely determined by the background $\Lambda$CDM-mimicking expansion history, the numerator contains the exact analytical growth rate and therefore encodes all modified-gravity effects. This separation also makes it straightforward to evaluate observational quantities characterizing the growth of cosmic structures.

\begin{figure}[H]
	\centering
	\subfigure[]{%
		\includegraphics[width=8cm,height=6cm]{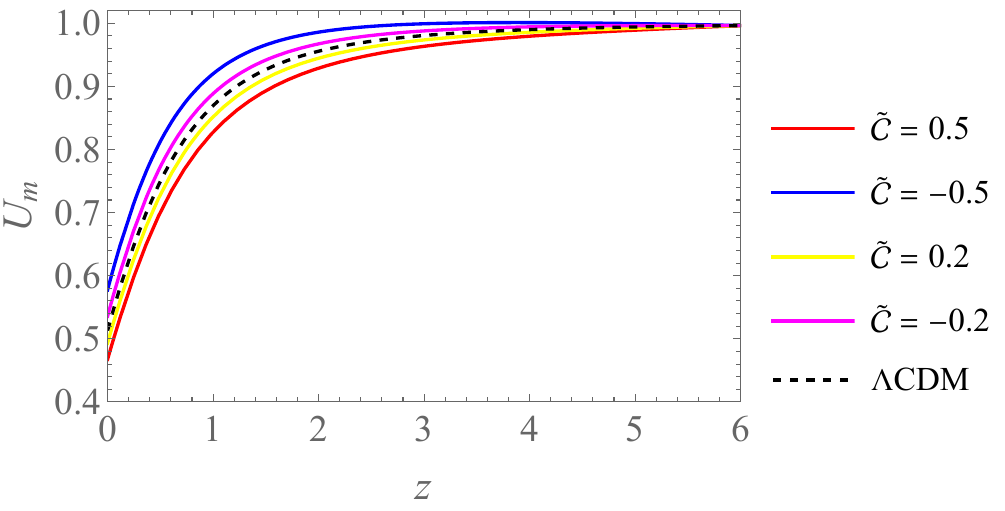}\label{fig:Um_z}}
	\quad
	\subfigure[]{%
		\includegraphics[width=8cm,height=6cm]{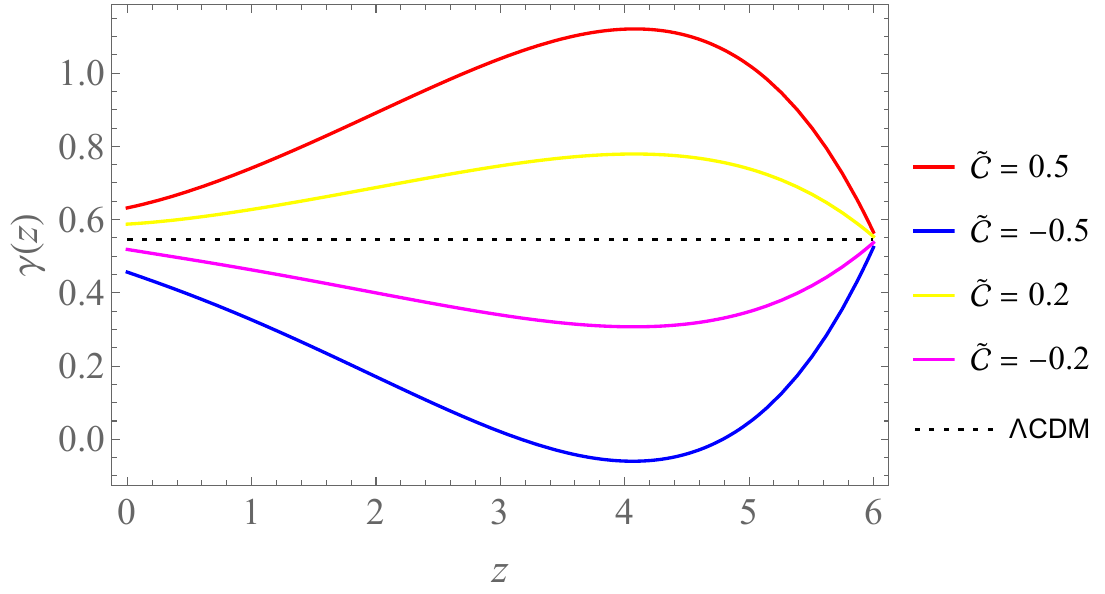}\label{fig:gamma_z}}
	\caption{The evolution of the matter perturbation growth function $U_m=\frac{\delta_m'}{\delta_m}$ (left panel) and the growth-index parameter (right panel) $\gamma$ in the redshift range $0<z<6$ for different $\Lambda$CDM-mimicking $f(Q)$ cosmologies, as specified by different values of $\tilde{\mathcal{C}}$ (Eq.\eqref{Omegavsq}). The evolutions are compared with that of the standard GR $\Lambda$CDM model, shown by the black dashed curve/line in both panels.}
    \label{fig:ptbn_soln}
\end{figure}

The analytical solution of $U_m(z)$ and $\gamma(z)$ for different $\Lambda$CDM-mimicking $f(Q)$ cosmologies, as specified by different values of the constant $\tilde{\mathcal{C}}$, is portrayed in Fig.\ref{fig:ptbn_soln} in the redshift range $0<z<6$. 
The perturbation evolution starts from a value that is close to GR at around $z\sim6$, which is expected because the integration constant $\mathcal{C}_1$ in the analytical solution of the perturbation is set to match GR when $q\lesssim0.5$. With time, the perturbation evolution for $\Lambda$CDM-mimicking $f(Q)$ cosmologies starts deviating from that of the GR-based $\Lambda$CDM model. Interestingly, Fig.\ref{fig:gamma_z} shows that even the growth index parameter, which is a characteristic of the perturbation evolution, even though it deviates significantly from the corresponding GR value $\sim6/11$ during the intermediate phase of the evolution, actually starts approaching the same asymptotically near the present epoch. This points towards a further degeneracy with the standard GR-based $\Lambda$CDM model, which goes beyond the background level. The somewhat unexpected result calls for a more detailed comparison at the perturbative level between the GR-based $\Lambda$CDM model and the $\Lambda$CDM mimicking $f(Q)$ models, something which we reserve for future work.

It can be seen from Fig.\ref{fig:ptbn_soln} that for all the $\Lambda$CDM-mimicking $f(Q)$ cosmological solutions considered, the growth function $U_m(z)$ is positive, and hence produces a real-valued growth index parameter $\gamma(z)$. Physically, this is a consistency check on the model because a positive $U_m$ implies the existence of a growing mode in matter perturbation, leading to astrophysical structure formation.

\section{Summary \& Discussion}\label{ss4}

The main focus of our paper has been to investigate $\Lambda$CDM-mimicking $f(Q)$ gravity, within the first connection branch, at both the background and linear perturbation levels. To this goal, we have employed the autonomous dynamical system approach. Since the form of such underlying $f(Q)$ theory can be exactly reconstructed, namely
$f(Q)=-2\Lambda+\alpha Q+\beta\sqrt{-Q}$, one can formulate a closed cosmological dynamical system for the $\Lambda$CDM-mimicking $f(Q)$ cosmology in two different ways. Firstly, one can form a closed system by specifying the functional form of the theory, which has been the subject matter of Sec.~\ref{ss2a}. Secondly, one can form a closed system by imposing  the underlying cosmology to be $\Lambda$CDM-mimicking, i.e., that it satisfies the condition $j(z)=1$, as discussed in Sec.~\ref{ss2b}. In both approaches, we could not only identify the fixed points, but also obtain analytical solutions and the corresponding phase trajectories.

Qualitatively, however, the two approaches describe the cosmological phase space in somewhat different manners. The phase portrait in the \emph{theory-driven closure} scheme,  Fig.~\ref{fig:phase_theo_G1}, shows how $\Omega=\Omega_m/f_Q$ evolves as the theory moves away from the GR limit, as also expressed by the analytical solution \eqref{Omega_in_m}. On the other hand, the phase portrait in the \emph{cosmography-driven closure} scheme,  Fig.~\ref{fig:phase_cosmo_G1}, shows how $\Omega$ evolves as the underlying cosmology gradually transits from one epoch to another, as expressed by the analytical solution \eqref{Omegavsq}. The interesting aspect is that the analytical solutions $\Omega(m)$ and $\Omega(q)$ (where $m=Qf_{QQ}/f_Q$ and $q$ is the deceleration parameter) can actually be related to each other through the parameter identifications discussed in Eq.~\eqref{one-to-one}. Hence, the two closure schemes provide complementary descriptions of the same cosmological problem and formulate the evolution in terms of different physical variables.

This correspondence  between the two approaches indicates that when it is not possible to reconstruct the $\Lambda$CDM-mimicking modified gravitational theory in a compact form, one can always bypass this problem by simply redesigning the dynamical system in such a way as to integrate the cosmographic condition $j=1$. This is important because not every family of modified gravity theories admits a compact functional form for the $\Lambda$CDM-mimicking theory. For example, there are three symmetric teleparallel $f(Q)$ connection branches in FLRW cosmology,  with the third  not admitting an explicit compact form for the functional  of the $\Lambda$CDM-mimicking $f(Q)$ \cite{Chakraborty:2025qlv}.  Similarly, in $f(R)$ gravity, the reconstructed $\Lambda$CDM-mimicking theory is expressed in terms of hypergeometric functions, making a direct construction of an autonomous dynamical system difficult \cite{Dunsby:2010wg,He:2012rf}: In this case, the \emph{cosmography-driven closure} approach has proven useful \cite{Chakraborty:2021jku,Chakraborty:2025lkz}.

Notably,  the first connection branch of $f(Q)$ gravity offers the special advantage of gauging away the connection function and reducing the field equations to the second order.  This simplicity has been advantageous for comparing the two closure formalisms and for establishing the correspondence between them.

The cosmography-driven closure scheme  also allows us to conduct further analytical analysis without resorting to numerical techniques. We have derived analytical expressions for the time evolution of key physical quantities. These encompass observationally constrainable parameters like $\kappa_{\rm eff}=1/f_Q$ and $\Omega_m$, alongside variables that capture deviations from General Relativity, such as $r=\frac{Q f_Q}{f}$ and $m=\frac{Qf_{QQ}}{f_Q}$ (with $\{r,m\}=\{1,0\}$ representing pure GR without a cosmological constant).  Furthermore, we have established the structural stability of the qualitative features of the $\Lambda$CDM-mimicking $f(Q)$ cosmological dynamics with respect to small deviations from the exact condition $j=1$ (Sect.~\ref{sec:robustness}), as well as the stability of $\Lambda$CDM-like and almost $\Lambda$CDM-like cosmological solutions against small homogeneous and isotropic perturbations (Sect.~\ref{sec:stability}).

Perhaps unexpectedly, the cosmographic closure also allows us to obtain analytical results for the linear matter perturbations. In particular, we could obtain analytical solutions for the growth function $U_m$ and the growth-index parameter $\gamma$ for the $\Lambda$CDM-mimicking $f(Q)$ scenario in the first connection branch. The evolution of $\gamma$ shown in Fig.~\ref{fig:ptbn_soln} for different choices of the boundary condition $\Omega_0$ in the $\Lambda$CDM-mimicking $f(Q)$ theory shows that the matter perturbations remain close to their GR-$\Lambda$CDM counterpart during both the matter-dominated and late-time epochs, with the largest deviations occurring during the transition between these two epochs.

Both $f(R)$ gravity and the $f(Q)$ gravity in the first connection branch have only metric as a dynamical degree of freedom, the former leading to fourth-order field equations, while the latter to second-order. Different aspects of $\Lambda$CDM-mimicking $f(R)$ gravity have been studied across various works utilizing the cosmographic closure approach \cite{Chakraborty:2021jku,MacDevette:2024wpg,Chakraborty:2025lkz,Samanta:2026pvb}. Phenomenologically, a comparative analysis of these two theories is in order,  given the shared aspects investigated in both frameworks. Without going into a detailed discussion, let us summarize below the obtained results pointwise:

\begin{itemize}
 \item \textbf{Phase portrait:} In the $\Lambda$CDM-mimicking $f(Q)$ cosmology considered here, the matter-dominated fixed point $B_3$ is a past attractor, whereas the corresponding point $P_3$ in $\Lambda$CDM-mimicking $f(R)$ gravity is a saddle \cite{Chakraborty:2021jku}. This difference leads to a qualitatively distinct structure of the cosmological trajectories in the two theories.

\item \textbf{GR as a past attractor:} The matter-dominated past attractor $B_3$ obtained in the cosmography-driven closure is essentially the same as the past attractor $A_2$ in the theory-driven closure. Since $A_2$ lies on the line $m=0$ (see Fig.~\ref{fig:phase_theo_G1a}), the GR limit acts as a cosmological past attractor for the $\Lambda$CDM-mimicking $f(Q)$ cosmologies considered here. This is also evident from Fig.~\ref{fig:GR_past_attractor}. The same conclusion does not hold for $\Lambda$CDM-mimicking $f(R)$ gravity, for which the matter-dominated fixed point is not a past attractor \cite{Chakraborty:2021jku,Chakraborty:2025lkz}.

\item \textbf{Robustness and perturbations:} We have found that the qualitative features of the $\Lambda$CDM-mimicking $f(Q)$ cosmology are preserved for small deviations from the cosmographic condition $j=1$, and that the corresponding cosmological solutions are stable against small homogeneous and isotropic perturbations. These features are not shared by the likewise $\Lambda$CDM-mimicking $f(R)$ models \cite{Chakraborty:2025lkz}. At the perturbative level, however, both theories display a similar qualitative evolution of the growth-index parameter: it approaches the $\Lambda$CDM value deep in the matter-dominated epoch, deviates from it during the transition to accelerated expansion, and approaches the $\Lambda$CDM value again near the present epoch \cite{Samanta:2026pvb}; we note that the present epoch behavior arises naturally as a consequence of the cosmological dynamics rather than constituting an ad-hoc imposition.

\end{itemize}
Let us conclude our paper with a few words on the possibility of addressing cosmological tensions.

 $\Lambda$CDM-mimicking cosmologies have been proposed as potential remedies to known observational tensions because they distinguish between kinematical and dynamical aspects of the cosmic evolution. The basic idea is that while the redshift dependence of the expansion rate is still consistent with the Hubble diagram as obtained from, say, the cosmic chronometer dataset, the underlying model can still admit some freedom in its free parameters at the perturbative level. A minimal example within GR is provided by $\Lambda$CDM-mimicking coupled dark matter-dark energy models \cite{Chakraborty:2022evc}. 
Even though they are indistinguishable from $\Lambda$CDM at the background level, the very presence of the dark sector coupling would introduce distinctive features at the perturbative level: This freedom might be utilized in our favour to address tensions at the perturbation level, namely the $f\sigma_8$ tension.

In the first connection branch of $f(Q)$ cosmology, the $\sqrt{-Q}$ term entering the functional $f(Q)$ (\ref{flcdmG1}) is precisely known to be consistent with a $\Lambda$CDM kinematics. It has been shown that by tuning the coefficient of the $\sqrt{-Q}$ term, this model can fulfill astrophysical constraints on the effective gravitational coupling \cite{Chakraborty:2025qlv,Dutta:2025fqw},  Planck data and redshift space distortion surveys \cite{Atayde:2021pgb,Li:2025msm}.  However, this very term distinguishes it from the GR-$\Lambda$CDM model at the perturbative level, and we can therefore exploit this distinction  to our advantage.

Previous works along this direction, e.g., \cite{Atayde:2021pgb,Li:2025msm}, explicitly rely on numerical analysis. Our analytical solutions, presented in Sect.\ref{sec:ptbn}, make it easier to compute $f\sigma_8(z)$ and the diagnostic
\beq
\Delta(f\sigma_8)(z)=f\sigma_8^{f(Q)}(z)-f\sigma_8^{\Lambda{\rm CDM}}(z)\,,
\eeq
without resorting to numerical integration. As a result, we can have a fully analytical template for confronting $\Lambda$CDM-mimicking $f(Q)$ gravity with present and forthcoming redshift-space distortion surveys.

It is also worth recalling that $f(Q)$ cosmology may jointly address the Hubble and $\sigma_8$ tensions \cite{Boiza:2025xpn}. Moreover, our Eq.~(\ref{finalgamma}) for the relative growth rate is independent of $H_0$. This potentially allows the growth sector to be tested without directly entangling it with the value adopted for the present-day Hubble constant \cite{Forconi:2025cwp}. This property provides a promising starting point for a dedicated observational analysis.

More broadly, the results of this work show that the cosmographic perspective can provide a useful analytical route to the dynamics and perturbations of $\Lambda$CDM-mimicking modified gravity even when the corresponding gravitational theory is not available in a compact closed form. The first connection branch of $f(Q)$ gravity provides a particularly transparent example in which the theory-driven and cosmography-driven descriptions can be compared directly. The analytical treatment developed here, therefore, provides a framework that can be extended to other connection branches and, more generally, to other modified-gravity theories.

\section*{Acknowledgments}
DG is a member of the GNFM working group of the Italian INDAM. JD thank IUCAA, Pune, India for Visiting Associateship program. 

\appendix

\section{$f(Q)\propto\sqrt{-Q}$ as an effective boundary term in the FLRW minisuperspace}\label{appA}

We start from the point-like minisuperspace action for $f(Q)$ gravity in the spatially flat FLRW case ($k=0$) for the trivial connection branch $\Gamma_1$ \cite{Bajardi:2023vcc,Paliathanasis:2023pqp}:
\begin{equation}
S = \int dt L = \int dt a^3 \big( f - Q f_Q \big) - 6 a \dot{a}^2 f_Q\,.
\label{eq:L_general}
\end{equation}
Applying it to the special case $f(Q) = \beta \sqrt{-Q}$, the minisuperspace Lagrangian can be written as
\begin{align}
L(a,\dot{a},Q) &= a^3 \left(\frac{\beta}{2}\sqrt{-Q}\right) - 6 a \dot{a}^2 \left(-\frac{\beta}{2\sqrt{-Q}}\right) \\
&= \frac{\beta}{2} a^3 \sqrt{-Q} + \frac{3\beta}{\sqrt{-Q}} a \dot{a}^2.
\label{eq:L_subs}
\end{align}
Substituting now $Q = -6H^2 = -6\frac{\dot{a}^2}{a^2}$, and simplifying, one arrives at
\begin{align}
L(a,\dot{a}) &= \left( \frac{\beta \sqrt{6}}{2} + \frac{3\beta}{\sqrt{6}} \right) a^2 \dot{a} \\
&= \beta \sqrt{6}\, a^2 \dot{a} \\
& = \beta \sqrt{6}\, \frac{d}{dt} \left(\frac{1}{3} a^3 \right)
\end{align}
Thus the action reduces to a \emph{pure boundary term},
\begin{equation}
S = \beta \sqrt{6}\, \frac{a^3}{3}\Big|_{\text{boundary}},
\end{equation}
implying that the choice $f(Q) \propto \sqrt{-Q}$ does not yield any dynamical field equations in the minisuperspace formalism.

\bibliography{refs}

@article{Dutta:2025fqw,
    author = "Dutta, Jibitesh and Khyllep, Wompherdeiki and Chakraborty, Saikat and Gregoris, Daniele and Karwan, Khamphee",
    title = "{A unified dynamical systems framework for cosmology in f(Q) gravity: generic features across the connection branches}",
    eprint = "2508.09530",
    archivePrefix = "arXiv",
    primaryClass = "gr-qc",
    doi = "10.1140/epjc/s10052-025-15151-4",
    journal = "Eur. Phys. J. C",
    volume = "85",
    number = "12",
    pages = "1425",
    year = "2025"
}

@article{Hu:2023juh,
    author = "Hu, Yu-Min and Zhao, Yaqi and Ren, Xin and Wang, Bo and Saridakis, Emmanuel N. and Cai, Yi-Fu",
    title = "{The effective field theory approach to the strong coupling issue in f(T) gravity}",
    eprint = "2302.03545",
    archivePrefix = "arXiv",
    primaryClass = "gr-qc",
    doi = "10.1088/1475-7516/2023/07/060",
    journal = "JCAP",
    volume = "07",
    pages = "060",
    year = "2023"
}

@article{Bahamonde:2021gfp,
    author = "Bahamonde, Sebastian and Dialektopoulos, Konstantinos F. and Escamilla-Rivera, Celia and Farrugia, Gabriel and Gakis, Viktor and Hendry, Martin and Hohmann, Manuel and Levi Said, Jackson and Mifsud, Jurgen and Di Valentino, Eleonora",
    title = "{Teleparallel gravity: from theory to cosmology}",
    eprint = "2106.13793",
    archivePrefix = "arXiv",
    primaryClass = "gr-qc",
    doi = "10.1088/1361-6633/ac9cef",
    journal = "Rept. Prog. Phys.",
    volume = "86",
    number = "2",
    pages = "026901",
    year = "2023"
}

@article{BeltranJimenez:2017tkd,
    author = "Beltr{\'a}n Jim{\'e}nez, Jose and Heisenberg, Lavinia and Koivisto, Tomi",
    title = "{Coincident General Relativity}",
    eprint = "1710.03116",
    archivePrefix = "arXiv",
    primaryClass = "gr-qc",
    reportNumber = "NORDITA-2017-100, IFT-UAM/CSIC-17-093, ITS-ETH-2017-10",
    doi = "10.1103/PhysRevD.98.044048",
    journal = "Phys. Rev. D",
    volume = "98",
    number = "4",
    pages = "044048",
    year = "2018"
}

@article{DAmbrosio:2021pnd,
    author = "D'Ambrosio, Fabio and Heisenberg, Lavinia and Kuhn, Simon",
    title = "{Revisiting cosmologies in teleparallelism}",
    eprint = "2109.04209",
    archivePrefix = "arXiv",
    primaryClass = "gr-qc",
    doi = "10.1088/1361-6382/ac3f99",
    journal = "Class. Quant. Grav.",
    volume = "39",
    number = "2",
    pages = "025013",
    year = "2022"
}

@article{Heisenberg:2023lru,
    author = "Heisenberg, Lavinia",
    title = "{Review on f(Q) gravity}",
    eprint = "2309.15958",
    archivePrefix = "arXiv",
    primaryClass = "gr-qc",
    doi = "10.1016/j.physrep.2024.02.001",
    journal = "Phys. Rept.",
    volume = "1066",
    pages = "1--78",
    year = "2024"
}

@article{Zhao:2024kri,
    author = "Zhao, Dehao",
    title = "{Conformal transformation of f(Q) gravity and its cosmological perturbations}",
    eprint = "2404.16299",
    archivePrefix = "arXiv",
    primaryClass = "gr-qc",
    doi = "10.1103/PhysRevD.110.124034",
    journal = "Phys. Rev. D",
    volume = "110",
    number = "12",
    pages = "124034",
    year = "2024"
}

@article{Bahamonde:2017ize,
    author = {Bahamonde, Sebastian and B{\"o}hmer, Christian G. and Carloni, Sante and Copeland, Edmund J. and Fang, Wei and Tamanini, Nicola},
    title = "{Dynamical systems applied to cosmology: dark energy and modified gravity}",
    eprint = "1712.03107",
    archivePrefix = "arXiv",
    primaryClass = "gr-qc",
    doi = "10.1016/j.physrep.2018.09.001",
    journal = "Phys. Rept.",
    volume = "775-777",
    pages = "1--122",
    year = "2018"
}

@article{Atayde:2026upv,
    author = "Atayde, Lu{\'\i}s and Marques Nunes, Sim{\~a}o and Frusciante, Noemi",
    title = "{Inverse nonmetricity in f(Q) gravity: Cosmology and observational constraints}",
    eprint = "2603.27428",
    archivePrefix = "arXiv",
    primaryClass = "astro-ph.CO",
    doi = "10.1103/ts4w-xn8v",
    journal = "Phys. Rev. D",
    volume = "114",
    number = "2",
    pages = "024056",
    year = "2026"
}

@article{Atayde:2021pgb,
    author = "Atayde, Lu{\'\i}s and Frusciante, Noemi",
    title = "{Can $f(Q)$ gravity challenge $\Lambda$CDM?}",
    eprint = "2108.10832",
    archivePrefix = "arXiv",
    primaryClass = "astro-ph.CO",
    doi = "10.1103/PhysRevD.104.064052",
    journal = "Phys. Rev. D",
    volume = "104",
    number = "6",
    pages = "064052",
    year = "2021"
}

@article{MacDevette:2024wpg,
    author = "MacDevette, Kelly and Worsley, Jess and Dunsby, Peter and Chakraborty, Saikat",
    title = "{A model-independent approach to the study of structure growth in f(R) gravity}",
    eprint = "2408.03998",
    archivePrefix = "arXiv",
    primaryClass = "gr-qc",
    doi = "10.1093/mnras/staf168",
    journal = "Mon. Not. Roy. Astron. Soc.",
    volume = "537",
    number = "3",
    pages = "2471--2495",
    year = "2025"
}

@article{Capozziello:2024lsz,
    author = "Capozziello, Salvatore and Shokri, Mehdi",
    title = "{Comparing inflationary models in extended Metric-Affine theories of gravity}",
    eprint = "2408.17415",
    archivePrefix = "arXiv",
    primaryClass = "gr-qc",
    doi = "10.1016/j.dark.2024.101698",
    journal = "Phys. Dark Univ.",
    volume = "46",
    pages = "101698",
    year = "2024"
}

@article{Li:2025msm,
    author = "Li, Chunyu and Ren, Xin and Yang, Yuhang and Saridakis, Emmanuel N. and Cai, Yi-Fu",
    title = "{Decoupling perturbations from background in f(Q) gravity: the square-root correction and the impact on the {\ensuremath{\sigma}} $_{8}$ tension}",
    eprint = "2512.16551",
    archivePrefix = "arXiv",
    primaryClass = "astro-ph.CO",
    doi = "10.1088/1475-7516/2026/07/027",
    journal = "JCAP",
    volume = "07",
    pages = "027",
    year = "2026"
}

@article{Lin:2021uqa,
    author = "Lin, Rui-Hui and Zhai, Xiang-Hua",
    title = "{Spherically symmetric configuration in $f(Q)$ gravity}",
    eprint = "2105.01484",
    archivePrefix = "arXiv",
    primaryClass = "gr-qc",
    doi = "10.1103/PhysRevD.103.124001",
    journal = "Phys. Rev. D",
    volume = "103",
    number = "12",
    pages = "124001",
    year = "2021",
    note = "[Erratum: Phys.Rev.D 106, 069902 (2022)]"
}

@article{Nashed:2025bxv,
    author = "Nashed, G. G. L. and Saridakis, Emmanuel N.",
    title = "{Three-dimensional charged black holes in f(Q) gravity}",
    eprint = "2506.10046",
    archivePrefix = "arXiv",
    primaryClass = "gr-qc",
    doi = "10.1142/s0218271826500185",
    journal = "Int. J. Mod. Phys. D",
    volume = "35",
    number = "08",
    pages = "2650018",
    year = "2026"
}

@article{Capozziello:2022tvv,
    author = "Capozziello, Salvatore and Shokri, Mehdi",
    title = "{Slow-roll inflation in f(Q) non-metric gravity}",
    eprint = "2209.06670",
    archivePrefix = "arXiv",
    primaryClass = "gr-qc",
    doi = "10.1016/j.dark.2022.101113",
    journal = "Phys. Dark Univ.",
    volume = "37",
    pages = "101113",
    year = "2022"
}

@article{Frusciante:2021sio,
    author = "Frusciante, Noemi",
    title = "{Signatures of $f(Q)$-gravity in cosmology}",
    eprint = "2101.09242",
    archivePrefix = "arXiv",
    primaryClass = "astro-ph.CO",
    doi = "10.1103/PhysRevD.103.044021",
    journal = "Phys. Rev. D",
    volume = "103",
    number = "4",
    pages = "044021",
    year = "2021"
}

@article{Chaudhary:2025bfs,
    author = "Chaudhary, Himanshu and Sharma, Vipin Kumar and Capozziello, Salvatore and Mustafa, G.",
    title = "{Probing Departures from {\ensuremath{\Lambda}}CDM by Late-time Datasets}",
    eprint = "2510.08339",
    archivePrefix = "arXiv",
    primaryClass = "astro-ph.CO",
    doi = "10.3847/1538-4365/ae4b3f",
    journal = "Astrophys. J. Suppl.",
    volume = "283",
    number = "2",
    pages = "73",
    year = "2026"
}

@article{Forconi:2025cwp,
    author = "Forconi, Matteo and Favale, Arianna and G{\'o}mez-Valent, Adri{\`a}",
    title = "{Illustrating the consequences of a misuse of {\ensuremath{\sigma}}8 in cosmology}",
    eprint = "2501.11571",
    archivePrefix = "arXiv",
    primaryClass = "astro-ph.CO",
    doi = "10.1103/rpf5-ldks",
    journal = "Phys. Rev. D",
    volume = "112",
    number = "2",
    pages = "023517",
    year = "2025"
}

@article{Capozziello:2025qmh,
    author = "Capozziello, Salvatore and Chaudhary, Himanshu and Harko, Tiberiu and Mustafa, Ghulam",
    title = "{Is dark energy dynamical in the DESI era? A critical review}",
    eprint = "2512.10585",
    archivePrefix = "arXiv",
    primaryClass = "astro-ph.CO",
    doi = "10.1016/j.dark.2025.102196",
    journal = "Phys. Dark Univ.",
    volume = "51",
    pages = "102196",
    year = "2026"
}

@article{Adi:2025hyj,
    author = "Adi, Tal",
    title = "{Lowering the horizon on Dark Energy: A late-time response to early solutions for the Hubble tension}",
    eprint = "2509.12331",
    archivePrefix = "arXiv",
    primaryClass = "astro-ph.CO",
    doi = "10.1088/1475-7516/2026/03/015",
    journal = "JCAP",
    volume = "03",
    pages = "015",
    year = "2026"
}

@article{Paliathanasis:2025jlw,
    author = "Paliathanasis, Andronikos",
    title = "{The {\textquotedblleft}Telephone Game{\textquotedblright} effect in modern gravity research}",
    eprint = "2508.16682",
    archivePrefix = "arXiv",
    primaryClass = "gr-qc",
    doi = "10.1140/epjc/s10052-025-14553-8",
    journal = "Eur. Phys. J. C",
    volume = "85",
    number = "8",
    pages = "822",
    year = "2025"
}

@article{bressoud2011historical,
  author    = {Bressoud, David M.},
  title     = {Historical Reflections on Teaching the Fundamental Theorem of Integral Calculus},
  journal   = {The American Mathematical Monthly},
  volume    = {118},
  number    = {2},
  pages     = {99--115},
  year      = {2011},
  publisher = {Taylor \& Francis},
  doi       = {10.4169/amer.math.monthly.118.02.099},
  url       = {https://doi.org/10.4169/amer.math.monthly.118.02.099}
}

@article{Capozziello:2026rrc,
    author = "Capozziello, Salvatore and Capriolo, Maurizio and Lambiase, Gaetano",
    title = "{Gravitational energy-momentum pseudotensor in $f(Q)$ nonmetric gravity}",
    eprint = "2601.12088",
    archivePrefix = "arXiv",
    primaryClass = "gr-qc",
    doi = "10.1103/1kvt-vvcf",
    journal = "Phys. Rev. D",
    volume = "113",
    number = "4",
    pages = "044017",
    year = "2026"
}

@article{Murtaza:2025gra,
    author = "Murtaza, Ghulam and De, Avik and Paliathanasis, Andronikos and Loo, Tee-How",
    title = "{Can an extra degree of freedom in scalar-tensor non-metricity gravity account for the evolution of the Universe?}",
    eprint = "2506.17099",
    archivePrefix = "arXiv",
    primaryClass = "gr-qc",
    doi = "10.1088/1361-6382/ae0404",
    journal = "Class. Quant. Grav.",
    volume = "42",
    number = "19",
    pages = "195004",
    year = "2025"
}

@article{DESI:2025zgx,
    author = "Abdul Karim, M. and others",
    collaboration = "DESI",
    title = "{DESI DR2 results. II. Measurements of baryon acoustic oscillations and cosmological constraints}",
    eprint = "2503.14738",
    archivePrefix = "arXiv",
    primaryClass = "astro-ph.CO",
    reportNumber = "FERMILAB-PUB-25-0169-PPD",
    doi = "10.1103/tr6y-kpc6",
    journal = "Phys. Rev. D",
    volume = "112",
    number = "8",
    pages = "083515",
    year = "2025"
}

@article{Chakraborty:2025qlv,
    author = "Chakraborty, Saikat and Dutta, Jibitesh and Gregoris, Daniele and Karwan, Khamphee and Khyllep, Wompherdeiki",
    title = "{Reproducing {\ensuremath{\Lambda}}CDM-like solutions in f(Q) gravity: a~comprehensive study across all connection branches}",
    eprint = "2501.15159",
    archivePrefix = "arXiv",
    primaryClass = "gr-qc",
    doi = "10.1088/1475-7516/2025/05/098",
    journal = "JCAP",
    volume = "05",
    pages = "098",
    year = "2025"
}

@article{Jesus:2022xwb,
    author = "Jesus, J. F. and Benndorf, D. and Escobal, A. A. and Pereira, S. H.",
    title = "{From Hubble to snap parameters: a Gaussian process reconstruction}",
    eprint = "2212.12346",
    archivePrefix = "arXiv",
    primaryClass = "astro-ph.CO",
    doi = "10.1093/mnras/stae120",
    journal = "Mon. Not. Roy. Astron. Soc.",
    volume = "528",
    number = "2",
    pages = "1573--1581",
    year = "2024"
}

@article{Murtaza:2025gme,
    author = "Murtaza, Ghulam and Chakraborty, Saikat and De, Avik",
    title = "{A generic dynamical system formulation for Bianchi-I cosmology with isotropic fluid in f(Q) gravity}",
    eprint = "2504.21757",
    archivePrefix = "arXiv",
    primaryClass = "gr-qc",
    doi = "10.1088/1475-7516/2025/08/093",
    journal = "JCAP",
    volume = "08",
    pages = "093",
    year = "2025"
}

@article{BeltranJimenez:2019tme,
    author = "Beltr{\'a}n Jim{\'e}nez, Jose and Heisenberg, Lavinia and Koivisto, Tomi Sebastian and Pekar, Simon",
    title = "{Cosmology in $f(Q)$ geometry}",
    eprint = "1906.10027",
    archivePrefix = "arXiv",
    primaryClass = "gr-qc",
    doi = "10.1103/PhysRevD.101.103507",
    journal = "Phys. Rev. D",
    volume = "101",
    number = "10",
    pages = "103507",
    year = "2020"
}

@article{Visser:2003vq,
    author = "Visser, Matt",
    title = "{Jerk and the cosmological equation of state}",
    eprint = "gr-qc/0309109",
    archivePrefix = "arXiv",
    doi = "10.1088/0264-9381/21/11/006",
    journal = "Class. Quant. Grav.",
    volume = "21",
    pages = "2603--2616",
    year = "2004"
}

@article{Dunajski:2008tg,
    author = "Dunajski, Maciej and Gibbons, Gary",
    title = "{Cosmic Jerk, Snap and Beyond}",
    eprint = "0807.0207",
    archivePrefix = "arXiv",
    primaryClass = "gr-qc",
    reportNumber = "DAMTP-2008-58",
    doi = "10.1088/0264-9381/25/23/235012",
    journal = "Class. Quant. Grav.",
    volume = "25",
    pages = "235012",
    year = "2008"
}

@article{Capozziello:2019cav,
    author = "Capozziello, Salvatore and D'Agostino, Rocco and Luongo, Orlando",
    title = "{Extended Gravity Cosmography}",
    eprint = "1904.01427",
    archivePrefix = "arXiv",
    primaryClass = "gr-qc",
    doi = "10.1142/S0218271819300167",
    journal = "Int. J. Mod. Phys. D",
    volume = "28",
    number = "10",
    pages = "1930016",
    year = "2019"
}

@article{Chakraborty:2022evc,
    author = "Chakraborty, Saikat and Gregoris, Daniele and Mishra, B.",
    title = "{On the uniqueness of {\ensuremath{\Lambda}}CDM-like evolution for homogeneous and isotropic cosmology in General Relativity}",
    eprint = "2208.04596",
    archivePrefix = "arXiv",
    primaryClass = "gr-qc",
    doi = "10.1016/j.physletb.2023.137962",
    journal = "Phys. Lett. B",
    volume = "842",
    pages = "137962",
    year = "2023"
}

@article{Mukherjee:2016shl,
    author = "Mukherjee, Ankan and Banerjee, Narayan",
    title = "{In search of the dark matter dark energy interaction: a kinematic approach}",
    eprint = "1610.04419",
    archivePrefix = "arXiv",
    primaryClass = "astro-ph.CO",
    doi = "10.1088/1361-6382/aa54c8",
    journal = "Class. Quant. Grav.",
    volume = "34",
    number = "3",
    pages = "035016",
    year = "2017"
}

@article{Bernal:2016gxb,
    author = "Bernal, Jose Luis and Verde, Licia and Riess, Adam G.",
    title = "{The trouble with $H_0$}",
    eprint = "1607.05617",
    archivePrefix = "arXiv",
    primaryClass = "astro-ph.CO",
    doi = "10.1088/1475-7516/2016/10/019",
    journal = "JCAP",
    volume = "10",
    pages = "019",
    year = "2016"
}

@article{Mukherjee:2020ytg,
    author = "Mukherjee, Purba and Banerjee, Narayan",
    title = "{Non-parametric reconstruction of the cosmological $jerk$ parameter}",
    eprint = "2007.10124",
    archivePrefix = "arXiv",
    primaryClass = "astro-ph.CO",
    doi = "10.1140/epjc/s10052-021-08830-5",
    journal = "Eur. Phys. J. C",
    volume = "81",
    number = "1",
    pages = "36",
    year = "2021"
}

@article{Jiang:2024xnu,
    author = "Jiang, Jun-Qian and Pedrotti, Davide and da Costa, Simony Santos and Vagnozzi, Sunny",
    title = "{Nonparametric late-time expansion history reconstruction and implications for the Hubble tension in light of recent DESI and type Ia supernovae data}",
    eprint = "2408.02365",
    archivePrefix = "arXiv",
    primaryClass = "astro-ph.CO",
    doi = "10.1103/PhysRevD.110.123519",
    journal = "Phys. Rev. D",
    volume = "110",
    number = "12",
    pages = "123519",
    year = "2024"
}

@article{Gao:2025ozb,
    author = "Gao, Shengqing and Gao, Qing and Gong, Yungui and Lu, Xuchen",
    title = "{Null tests with Gaussian process}",
    eprint = "2503.15943",
    archivePrefix = "arXiv",
    primaryClass = "astro-ph.CO",
    doi = "10.1007/s11433-025-2682-1",
    journal = "Sci. China Phys. Mech. Astron.",
    volume = "68",
    number = "8",
    pages = "280408",
    year = "2025"
}

@article{Zhai:2013fxa,
    author = "Zhai, Zhong-Xu and Zhang, Ming-Jian and Zhang, Zhi-Song and Liu, Xian-Ming and Zhang, Tong-Jie",
    title = "{Reconstruction and constraining of the jerk parameter from OHD and SNe Ia observations}",
    eprint = "1303.1620",
    archivePrefix = "arXiv",
    primaryClass = "astro-ph.CO",
    doi = "10.1016/j.physletb.2013.10.020",
    journal = "Phys. Lett. B",
    volume = "727",
    pages = "8--20",
    year = "2013"
}

@article{Chakraborty:2021jku,
    author = "Chakraborty, Saikat and MacDevette, Kelly and Dunsby, Peter",
    title = "{A model independent approach to the study of $f(R)$ cosmologies with expansion histories close to $\Lambda$CDM}",
    eprint = "2103.02274",
    archivePrefix = "arXiv",
    primaryClass = "gr-qc",
    doi = "10.1103/PhysRevD.103.124040",
    journal = "Phys. Rev. D",
    volume = "103",
    number = "12",
    pages = "124040",
    year = "2021"
}

@article{Boehmer:2022wln,
    author = "Boehmer, Christian G. and Jensko, Erik and Lazkoz, Ruth",
    title = "{Cosmological dynamical systems in modified gravity}",
    eprint = "2201.09588",
    archivePrefix = "arXiv",
    primaryClass = "gr-qc",
    doi = "10.1140/epjc/s10052-022-10412-y",
    journal = "Eur. Phys. J. C",
    volume = "82",
    number = "6",
    pages = "500",
    year = "2022"
}

@article{Capozziello:2024vix,
    author = "Capozziello, Salvatore and Capriolo, Maurizio and Nojiri, Shin'ichi",
    title = "{Gravitational waves in f(Q) non-metric gravity via geodesic deviation}",
    eprint = "2401.06424",
    archivePrefix = "arXiv",
    primaryClass = "gr-qc",
    doi = "10.1016/j.physletb.2024.138510",
    journal = "Phys. Lett. B",
    volume = "850",
    pages = "138510",
    year = "2024"
}

@article{Hohmann:2017jao,
    author = "Hohmann, Manuel and Jarv, Laur and Ualikhanova, Ulbossyn",
    title = "{Dynamical systems approach and generic properties of $f(T)$ cosmology}",
    eprint = "1706.02376",
    archivePrefix = "arXiv",
    primaryClass = "gr-qc",
    doi = "10.1103/PhysRevD.96.043508",
    journal = "Phys. Rev. D",
    volume = "96",
    number = "4",
    pages = "043508",
    year = "2017"
}

@article{Jarv:2018bgs,
    author = {J{\"a}rv, Laur and R{\"u}nkla, Mihkel and Saal, Margus and Vilson, Ott},
    title = "{Nonmetricity formulation of general relativity and its scalar-tensor extension}",
    eprint = "1802.00492",
    archivePrefix = "arXiv",
    primaryClass = "gr-qc",
    doi = "10.1103/PhysRevD.97.124025",
    journal = "Phys. Rev. D",
    volume = "97",
    number = "12",
    pages = "124025",
    year = "2018"
}

@article{Bajardi:2023vcc,
    author = "Bajardi, Francesco and Capozziello, Salvatore",
    title = "{Minisuperspace quantum cosmology in f(Q) gravity}",
    eprint = "2305.00318",
    archivePrefix = "arXiv",
    primaryClass = "gr-qc",
    doi = "10.1140/epjc/s10052-023-11703-8",
    journal = "Eur. Phys. J. C",
    volume = "83",
    number = "6",
    pages = "531",
    year = "2023"
}

@article{Paliathanasis:2023pqp,
    author = "Paliathanasis, A. and Dimakis, N. and Christodoulakis, T.",
    title = "{Minisuperspace description of f(Q)-cosmology}",
    eprint = "2308.15207",
    archivePrefix = "arXiv",
    primaryClass = "gr-qc",
    doi = "10.1016/j.dark.2023.101410",
    journal = "Phys. Dark Univ.",
    volume = "43",
    pages = "101410",
    year = "2024"
}

@article{Albuquerque:2022eac,
    author = "Albuquerque, In{\^e}s S. and Frusciante, Noemi",
    title = "{A designer approach to f(Q) gravity and cosmological implications}",
    eprint = "2202.04637",
    archivePrefix = "arXiv",
    primaryClass = "astro-ph.CO",
    doi = "10.1016/j.dark.2022.100980",
    journal = "Phys. Dark Univ.",
    volume = "35",
    pages = "100980",
    year = "2022"
}

@article{Boiza:2025xpn,
    author = "Boiza, Carlos G. and Petronikolou, Maria and Bouhmadi-L{\'o}pez, Mariam and Saridakis, Emmanuel N.",
    title = "{Addressing H $_{0}$ and S $_{8}$ tensions within f(Q) cosmology}",
    eprint = "2505.18264",
    archivePrefix = "arXiv",
    primaryClass = "astro-ph.CO",
    doi = "10.1088/1475-7516/2025/12/011",
    journal = "JCAP",
    volume = "12",
    pages = "011",
    year = "2025"
}

@article{Rodrigues:2025tfg,
    author = "Rodrigues, Gabriel and de Souza, Rayff and Alcaniz, Jailson",
    title = "{Cosmography with DESI DR2 and SN data}",
    eprint = "2506.22373",
    archivePrefix = "arXiv",
    primaryClass = "astro-ph.CO",
    doi = "10.1103/fh98-1b3d",
    journal = "Phys. Rev. D",
    volume = "112",
    number = "10",
    pages = "103519",
    year = "2025"
}

@article{Dunsby:2010wg,
    author = "Dunsby, Peter K. S. and Elizalde, Emilo and Goswami, Rituparno and Odintsov, Sergei and Gomez, Diego Saez",
    title = "{On the LCDM Universe in f(R) gravity}",
    eprint = "1005.2205",
    archivePrefix = "arXiv",
    primaryClass = "gr-qc",
    doi = "10.1103/PhysRevD.82.023519",
    journal = "Phys. Rev. D",
    volume = "82",
    pages = "023519",
    year = "2010"
}

@article{He:2012rf,
    author = "He, Jian-hua and Wang, Bin",
    title = "{Revisiting $f(R)$ gravity models that reproduce $\Lambda$CDM expansion}",
    eprint = "1208.1388",
    archivePrefix = "arXiv",
    primaryClass = "astro-ph.CO",
    doi = "10.1103/PhysRevD.87.023508",
    journal = "Phys. Rev. D",
    volume = "87",
    number = "2",
    pages = "023508",
    year = "2013"
}

@article{Ortiz-Banos:2021jgg,
    author = "Ortiz-Ba{\~n}os, Mar{\'\i}a and Bouhmadi-L{\'o}pez, Mariam and Lazkoz, Ruth and Salzano, Vincenzo",
    title = "{${\Lambda}$CDM suitably embedded in f(R) with a non-minimal coupling to matter}",
    eprint = "2103.01982",
    archivePrefix = "arXiv",
    primaryClass = "gr-qc",
    doi = "10.1140/epjc/s10052-021-09004-z",
    journal = "Eur. Phys. J. C",
    volume = "81",
    number = "3",
    pages = "237",
    year = "2021"
}

@article{Myrzakulov:2010gt,
    author = "Myrzakulov, Ratbay and Saez-Gomez, Diego and Tureanu, Anca",
    title = "{On the $\Lambda$CDM Universe in $f(G)$ gravity}",
    eprint = "1009.0902",
    archivePrefix = "arXiv",
    primaryClass = "gr-qc",
    doi = "10.1007/s10714-011-1149-y",
    journal = "Gen. Rel. Grav.",
    volume = "43",
    pages = "1671--1684",
    year = "2011"
}

@article{Elizalde:2010jx,
    author = "Elizalde, E. and Myrzakulov, R. and Obukhov, V. V. and Saez-Gomez, D.",
    title = "{LambdaCDM epoch reconstruction from F(R,G) and modified Gauss-Bonnet gravities}",
    eprint = "1001.3636",
    archivePrefix = "arXiv",
    primaryClass = "gr-qc",
    doi = "10.1088/0264-9381/27/9/095007",
    journal = "Class. Quant. Grav.",
    volume = "27",
    pages = "095007",
    year = "2010"
}

@article{Setare:2013xh,
    author = "Setare, M. R. and Mohammadipour, N.",
    title = "{Can $f(T)$ gravity theories mimic $\Lambda$CDM cosmic history}",
    eprint = "1301.4891",
    archivePrefix = "arXiv",
    primaryClass = "gr-qc",
    doi = "10.1088/1475-7516/2013/01/015",
    journal = "JCAP",
    volume = "01",
    pages = "015",
    year = "2013"
}

@article{Anagnostopoulos:2021ydo,
    author = "Anagnostopoulos, Fotios K. and Basilakos, Spyros and Saridakis, Emmanuel N.",
    title = "{First evidence that non-metricity f(Q) gravity could challenge {\ensuremath{\Lambda}}CDM}",
    eprint = "2104.15123",
    archivePrefix = "arXiv",
    primaryClass = "gr-qc",
    doi = "10.1016/j.physletb.2021.136634",
    journal = "Phys. Lett. B",
    volume = "822",
    pages = "136634",
    year = "2021"
}

@article{Chakraborty:2025lkz,
    author = "Chakraborty, Saikat and Burikham, Piyabut",
    title = "{Theory space and stability analysis of General Relativistic cosmological solutions in modified gravity}",
    eprint = "2509.15762",
    archivePrefix = "arXiv",
    primaryClass = "gr-qc",
    doi = "10.1140/epjc/s10052-026-15390-z",
    journal = "Eur. Phys. J. C",
    volume = "86",
    number = "3",
    pages = "280",
    year = "2026"
}

@article{Chakraborty:2025rvc,
    author = "Chakraborty, Saikat and Louw, Charlotte and Dunsby, Peter K. S. and MacDevette, Kelly and de la Cruz Dombriz, Alvaro",
    title = "{Dynamical dark energy in models with an evolution close to {\ensuremath{\Lambda}}CDM}",
    eprint = "2508.09813",
    archivePrefix = "arXiv",
    primaryClass = "gr-qc",
    doi = "10.1103/57rv-kvm6",
    journal = "Phys. Rev. D",
    volume = "112",
    number = "10",
    pages = "103518",
    year = "2025"
}

@article{Samanta:2026pvb,
    author = "Samanta, Apurba and Joshi, Bhuwan and Worsley, Jess and Dunsby, Peter and Chakraborty, Saikat",
    title = "{Constraining Scale-Dependent Growth in $f(R)$ Gravity with Future 21 cm Surveys; astro-ph.CO/2606.03729}",
    eprint = "2606.03729",
    archivePrefix = "arXiv",
    primaryClass = "astro-ph.CO",
    month = "6",
    year = "2026"
}

@article{Minucci:2024qrn,
    author = "Minucci, Marica and Panosso Macedo, Rodrigo",
    title = "{The confluent Heun functions in black hole perturbation theory: a spacetime interpretation}",
    eprint = "2411.19740",
    archivePrefix = "arXiv",
    primaryClass = "gr-qc",
    doi = "10.1007/s10714-025-03364-7",
    journal = "Gen. Rel. Grav.",
    volume = "57",
    number = "2",
    pages = "33",
    year = "2025"
}

@article{Chen:2023ese,
    author = "Chen, Changkai and Jing, Jiliang",
    title = "{Radiation fluxes of gravitational, electromagnetic, and scalar perturbations in type-D black holes: an exact approach}",
    eprint = "2307.14616",
    archivePrefix = "arXiv",
    primaryClass = "gr-qc",
    doi = "10.1088/1475-7516/2023/11/070",
    journal = "JCAP",
    volume = "11",
    pages = "070",
    year = "2023"
}

@article{Worsley:2026ijo,
    author = "Worsley, Jess and Chakraborty, Saikat and Dunsby, Peter",
    title = "{Beyond $j=1$: Observational Constraints on Almost-$\Lambda$CDM Cosmologies; astro-ph.CO/2607.20348}",
    eprint = "2607.20348",
    archivePrefix = "arXiv",
    primaryClass = "astro-ph.CO",
    month = "7",
    year = "2026"
}
\bibliographystyle{unsrt}

\end{document}